\documentclass[final,3p,times,twocolumn]{elsarticle}
\usepackage{atlasphysics}
\usepackage{subfigure}
\usepackage{mathrsfs}
\usepackage{extra_definitions}
\usepackage[hyperindex,breaklinks]{hyperref} 

\usepackage{lineno}

\usepackage{url}

\usepackage{preprintcover} 

\PreprintCoverPaperTitle{Measurement of the t-channel single top-quark production 
       cross section in pp collisions at $\sqrt{s} = 7$~TeV
       with the ATLAS detector}

\PreprintIdNumber{CERN-PH-EP-2012-082}

\PreprintCoverAbstract{We report a measurement of the cross section of single top-quark production 
in the $t$-channel using 1.04~fb$^{-1}$ of $pp$ collision data 
at $\sqrt{s}=7$~\tev\ recorded with the ATLAS detector at the LHC. 
Selected events feature one electron or muon, missing transverse momentum, 
and two or three jets, exactly one of them identified as originating 
from a $b$~quark.
The cross section is measured by fitting the distribution of 
a multivariate discriminant constructed with a neural network, yielding 
$\sigma_{t}= 83\pm 4\,(\mathrm{stat.})\,^{+20}_{-19}\,(\mathrm{syst.})\,\mathrm{pb}$, 
which is in good agreement with the prediction of the Standard Model. 
Using the ratio of the measured to the theoretically predicted cross section and assuming 
that the top-quark-related CKM matrix elements obey the relation
$|V_{tb}|\gg |V_{ts}|, |V_{td}|$, the coupling strength 
at the $W$-$t$-$b$ vertex is determined to be 
$|V_{tb}|=1.13^{+0.14}_{-0.13}$. If it is assumed that $|V_{tb}|\leq 1$
a lower limit of $|V_{tb}|>0.75$ is obtained at the 95\% confidence level.}

\PreprintJournalName{Physics Letters B 717 (2012) 330-350}

\begin{document}
\begin{frontmatter}
\title{\vspace{-2.0cm}Measurement of the t-channel single top-quark production 
       cross section in $pp$ collisions at $\sqrt{s} = 7$~\tev\
       with the ATLAS detector}

\author{The ATLAS Collaboration}
\address{}

\begin{abstract}
We report a measurement of the cross section of single top-quark production 
in the $t$-channel using 1.04~fb$^{-1}$ of $pp$ collision data 
at $\sqrt{s}=7$~\tev\ recorded with the ATLAS detector at the LHC. 
Selected events feature one electron or muon, missing transverse momentum, 
and two or three jets, exactly one of them identified as originating 
from a $b$~quark.
The cross section is measured by fitting the distribution of 
a multivariate discriminant constructed with a neural network, yielding 
$\sigma_{t}= 83\pm 4\,(\mathrm{stat.})\,^{+20}_{-19}\,(\mathrm{syst.})\,\mathrm{pb}$, 
which is in good agreement with the prediction of the Standard Model. 
Using the ratio of the measured to the theoretically predicted cross section and assuming 
that the top-quark-related CKM matrix elements obey the relation
$|V_{tb}|\gg |V_{ts}|, |V_{td}|$, the coupling strength 
at the $W$-$t$-$b$ vertex is determined to be 
$|V_{tb}|=1.13^{+0.14}_{-0.13}$. If it is assumed that $|V_{tb}|\leq 1$
a lower limit of $|V_{tb}|>0.75$ is obtained at the 95\% confidence level.
\end{abstract}

\end{frontmatter}

\section{Introduction}

At hadron colliders top quarks are predominantly produced in pairs (top-antitop) via the 
flavour-conserving strong interaction. Alternative production modes proceed via the
weak interaction involving a $W$-$t$-$b$ vertex, leading to a single top-quark intermediate 
state. Three subprocesses contribute to single top-quark production: 
the exchange of a
virtual $W$ boson in the $t$-channel, or in the $s$-channel, and the associated production
of a top quark and an on-shell $W$~boson. The process with the highest cross
section at the Tevatron and at the LHC is the $t$-channel mode $q+b\rightarrow q^\prime + t$.

In 2009, single top-quark production was observed by the CDF~\cite{Aaltonen:2009jj} and 
D\O~\cite{Abazov:2009ii} Collaborations 
based on analyses counting the $t$-channel and $s$-channel processes as signal.
The observation of the $t$-channel production mode has also been recently 
reported by D\O~\cite{Abazov:2011rz}, while the CMS Collaboration has published 
evidence of this process at the LHC~\cite{PhysRevLett.107.091802}. 

The single top-quark final state provides a direct probe of the $W$-$t$-$b$ coupling and is 
sensitive to many models of new physics~\cite{Tait:2000sh}.
The measurement of the production cross section constrains the absolute value of 
the quark-mixing matrix element 
$V_{tb}$ \cite{CKM1,CKM2} without assumptions about the number of quark generations
(see Ref.~\cite{PhysRevD.84.112001} for a recent measurement from the D\O~Collaboration). 
Alternatively, it allows the $b$-quark parton distribution function (PDF) to be measured.

At the LHC, colliding protons at $\sqrt{s}=7\tev$, the sum of $t$ and $\tbar$ cross sections
is predicted to be: $\sigma_t = 64.6^{+2.7}_{-2.0}\;$pb~\cite{Kidonakis:2011wy} 
for the leading $t$-channel process, $\sigma_{Wt} = 15.7\pm1.1\;$pb~\cite{Kidonakis:2010ux} 
for $Wt$ associated production, and 
$\sigma_s = 4.6\pm0.2\;$pb~\cite{Kidonakis:2010tc} for the $s$-channel. 
The analyses presented in this Letter consider only the $t$-channel process as signal,
while the other two single top-quark processes are treated as backgrounds,
assuming the Standard Model (SM) theoretical cross sections for these processes.

The $W$ boson from the top-quark decay is reconstructed in its leptonic decay
modes $e\nu$, $\mu\nu$ or $\tau\nu$, where the $\tau$ decays leptonically.
Thus, selected events contain one charged lepton candidate, 
$e$ or $\mu$; two or three hadronic high-$p_\mathrm{T}$ jets; and missing transverse momentum $\MET$.
Two jets are expected from the leading-order (LO) process, while a third jet may arise
from higher-order processes. 
Exactly one of the jets is required to be identified as originating from a $b$-quark.

The measurement of $\sigma_t$ is based on a fit to a multivariate discriminant 
constructed with a neural network (NN) to separate signal from background
and the result is cross-checked using a cut-based method,
which additionally provides a breakdown for the $t$ and $\bar{t}$
cross sections.

\section{Data and simulated event samples}
\label{sec:samples}
The analyses described in this Letter use proton-proton LHC collision data at a
centre-of-mass energy of 7~\tev\ collected with the ATLAS detector~\cite{ATL-2008-001} 
between March and June 2011.
The selected events were recorded based on single electron and muon triggers. Stringent 
detector and data quality requirements are applied, resulting in a data set 
corresponding to an integrated luminosity of $1.04\pm0.04$~fb$^{-1}$~\cite{Aad:2011dr,ATLAS-CONF-2011-116}.

Samples of simulated events for all three single top-quark processes are
produced with the \textsc{AcerMC} program (version 3.7)~\cite{Kersevan:2004yg}
using MRST 2007LO$^{*}$ parton distribution functions (PDFs)~\cite{Sherstnev:2007nd}.
The computation of the $t$-channel single top-quark process in \textsc{AcerMC}
incorporates the $q+b\rightarrow q^\prime+ t$ and $q+g \rightarrow q^\prime +t +\bar{b}$
subprocesses and features an automated procedure to remove the overlap in
phase space between them~\cite{Kersevan:2006fq}.
Samples of the top-quark pair ($t\bar{t}$) process are generated using 
\textsc{MC@NLO} (version 3.41)~\cite{SAMPLES-MCNLO},
with the CTEQ6.6 set of PDFs~\cite{cteq6l}.
The top-quark mass is assumed to be $172.5\gev$. 
Generator default values of 0.999105 and 0.999152 are used for $|V_{tb}|$
to produce the \textsc{AcerMC} and \textsc{MC@NLO} samples, respectively.
At higher orders in perturbation theory, interference effects between the 
single-top $Wt$~channel and $t\bar{t}$ processes occur, but are found to be 
small~\cite{Frixione:2008yi} and can therefore be safely neglected.
The \textsc{ALPGEN} leading-order generator (version 2.13)~\cite{SAMPLES-ALPGEN}
and the CTEQ6L1 set of PDFs~\cite{cteq6l} are used to generate $W+$jets,
$Wb\bar{b}$, $Wc\bar{c}$, $Wc$ and $Z$+jets events
with up to five additional partons.
To remove overlaps between the $n$ and $n+1$ parton samples
the ``MLM'' matching scheme~\cite{SAMPLES-ALPGEN} is used.
The double counting between the inclusive $W+n$ parton samples and
samples with associated heavy-quark pair-production is removed utilising
an overlap removal based on a $\Delta R=\sqrt{(\Delta\eta)^2+(\Delta\phi)^2}$ matching\footnote{
  ATLAS uses a right-handed coordinate system with its origin at the
  nominal interaction point in the centre of the detector and the $z$-axis along
  the beam direction. The $x$-axis points towards the centre of the LHC ring, the
  $y$-axis points upwards. The pseudorapidity is defined in terms of the polar angle
  $\theta$ as $\eta = -\ln \tan(\theta/2)$.}.
The diboson processes $WW$, $WZ$ and $ZZ$ are generated using 
\textsc{HERWIG} (version 6.5.20)~\cite{SAMPLES-HERWIG}. 
For all single top-quark samples the hadronisation is performed by 
\textsc{PYTHIA} (version 6.4.25)~\cite{pythia_sgtop}; 
in all other cases \textsc{HERWIG} in connection with 
the \textsc{JIMMY}~\cite{Butterworth:1996zw} underlying event model 
(version 4.31) is used. 
After the event generation, all samples are passed through 
the full simulation of the ATLAS detector~\cite{ATL-2010-005_sgtop} 
based on \textsc{GEANT4}~\cite{SAMPLES-G4} 
and are then reconstructed using the same procedure as for collision data. 
The simulation includes the effect of multiple $pp$ collisions per bunch crossing 
and is weighted to the same distribution as observed in the data with 5.6 interactions 
per bunch crossing on average.

\section{Object definition and event selection}
\label{sec:objsel}

Electron candidates are reconstructed offline using a cluster-based algorithm and are required 
to have $\ET >25\gev$ and $|\eta_\mathrm{cl}|<2.47$, where 
$\eta_\mathrm{cl}\,$ denotes the pseudorapidity of the calorimeter cluster.
Clusters in the transition regions between the calorimeter barrel and endcaps, 
corresponding to $1.37<|\eta_\mathrm{cl}|<1.52$, are ignored.
High-quality electron candidates are selected using a set of cuts~\cite{Aad:2011mk} which include stringent
requirements on the matching between the track and the calorimeter cluster. 
Electrons must also be isolated: the sum of the calorimeter transverse energy 
within a cone of radius $\Delta R=0.3$ 
(excluding the cells associated with the electron) must be less than 15\% of the electron 
\ET, and the \pT \ of all tracks within the same cone  
radius around the electron direction, again excluding the track associated to the electron, 
must be less than 10\% of the electron \ET.

Muon candidates are reconstructed by combining track segments found in the inner 
detector and in the muon spectrometer. We only consider muon candidates that have  
$\pT>25\gev$ and $|\eta|<2.5$. Selected muons must additionally satisfy a series of 
cuts on the number of track hits present in the various tracking 
sub-detectors~\cite{Aad:2011yb}. Muon candidates are required to be isolated
using the equivalent criteria as applied to electron candidates.

Jets are reconstructed using the anti-$k_{t}$ algorithm~\cite{CAC-0801} 
with a radius parameter of 0.4, using clusters of adjacent calorimeter cells~\cite{:2010wv} 
as inputs to the jet clustering. 
The response of the calorimeter is corrected by \pT- and $\eta$-dependent factors~\cite{ATLAS-JES-2011}, 
which are applied to each jet to provide an average energy scale correction.
Jets overlapping with selected electron candidates within $\Delta R<0.2$ are 
removed, as in these cases the jet and the electron are very likely to correspond 
to the same physics object. 
Only jets having $\pT >25\gev$ and $|\eta|<4.5$ are considered.
Jets originating from bottom quarks are tagged in the region $|\eta|<2.5$ by 
reconstructing secondary and tertiary vertices from the tracks associated with 
each jet and combining lifetime-related information with a NN~\cite{ATLAS-CONF-2011-102}.
A threshold is applied to the $b$-tagging algorithm output corresponding to a
$b$-tagging efficiency of about 57\% and a light-quark jet rejection factor 
(the reciprocal of the efficiency to $b$-tag light quarks) of about 520
for jets in $t\bar{t}$ events. The $\MET$ is calculated using clusters of adjacent calorimeter cells
and corrected for the presence of electrons, muons, and jets~\cite{ATLAS-MET}.

Events are selected if they contain at least one good primary vertex candidate~\cite{ATLAS-CONF-2012-042}
with a minimum of five associated tracks each with $\pt > 400\mev$. 
Events containing jets failing quality criteria~\cite{ATLAS-CONF-2010-038} are rejected.

The event selection requires exactly one charged lepton, $e$ or $\mu$, 
exactly two or three jets, and $\MET>25\gev$. 
A trigger matching requirement is applied where the lepton must lie within 
$\Delta R<0.15$ of its trigger-level object. 
Since the multijet background is difficult to model precisely, its 
contribution is additionally reduced through a requirement on the transverse mass of the 
lepton-$\MET$ system\footnote{
Defined as \\ $m_{\mathrm{T}}(W)=\sqrt{(\pT(\ell)+\MET)^2-(p_x(\ell)+E_x^\mathrm{miss})^2-(p_y(\ell)+E_y^\mathrm{miss})^2}$,
where $\ell$ denotes the lepton.}: 
$m_{\mathrm{T}}(W) > (60\gev - \MET)$~\cite{top2010}.

The following samples are defined for this analysis: a ``$b$-tagged sample'' with two
jets or three jets, exactly one of which is $b$-tagged, and a ``pretag sample''
with two or three jets, without making any $b$-tagging requirement. We also use a 
sample containing exactly one $b$-tagged jet to estimate the $W$+jets flavour composition.

\section{Background estimation}\label{sec:bgestimation}

A large background to the single top-quark final state comes from QCD-produced multijet 
events in which either one of the jets is misidentified as an isolated lepton or a non-prompt 
lepton (for example from a $b$-quark semileptonic decay) appears isolated.
Other significant backgrounds originate from $W$-boson production in association with jets and \ttbar~production. 
Smaller backgrounds come from $Z$+jets, $Wt$-channel and $s$-channel single top-quark production, and diboson production. 
These smaller backgrounds and the \ttbar~background are normalised to their theoretical predictions.  
For $Z$+jets background the inclusive cross sections are calculated to next-to-next-to-leading order (NNLO) with \textsc{FEWZ} (version from March 15, 2009)~\cite{FEWZ}.
The diboson cross sections are normalised to next-to-leading order theoretical calculations~\cite{Campbell2011bn}.
The \ttbar~cross section is normalised to the approximate NNLO-predicted value obtained using \textsc{HATHOR} (version 1.2)~\cite{hathor}.

The multijet background normalisation is obtained using 
a binned maximum-likelihood fit to the \MET~distribution in the data,
before the application of the \MET~cut, using a data-derived
template for the multijet background and templates from Monte-Carlo simulation 
for all other processes (top quark, $W/Z$+jets, dibosons).
The multijet template is created using collision events that are triggered 
by a single low-$\pt$ jet. Several prescaled trigger streams with different
$\pt$ thresholds are used for that purpose.
In the offline selection of these events the electron requirement is replaced 
by a jet requirement (jet-electron model). This jet must have 
$\pt > 25\gev$, 
the same acceptance in $|\eta|$ as the signal electron, and (80--95)\% of the 
jet energy deposited in the electromagnetic section of the calorimeter.
The last requirement ensures the orthogonality of the jet-electron data set
to the sample of events with electron candidates which feature an electromagnetic energy fraction
larger than 95\%. 
The jet must also contain at least four tracks, thus reducing the contribution from 
converted photons. When selecting the jet-electron sample, 
events containing electron or muon candidates in addition to the 
jet-electron are vetoed. 
The same model is also used in the muon channel.
A systematic uncertainty of 50\% on the multijet background rates was estimated
by studying the impact of pile-up events on the fit results and by
performing likelihood fits on the $m_{\mathrm{T}}(W)$ distribution. 
The jet-electron model is also used to
model the shape of kinematic distributions of the multijet background.

The kinematic distributions of the $W+$jets background, which comprises 
contributions from $W$+heavy flavour jets ($Wb\bar{b}$+jets, $Wc\bar{c}$+jets and $Wc$+jets) and 
$W$+light jets, are taken from samples of simulated events,
while the normalisation of the flavour composition is derived from data.
The NN analysis simultaneously determines the normalisation of the $W+$light jets and 
$W+$ heavy flavour processes when fitting the NN discriminant distribution 
to measure the $t$-channel single top-quark rate.
The cut-based analysis derives normalisation factors for the $W+$jets processes using the 
event yields in the 1-jet $b$-tagged, 2-jet pretag, and 2-jet $b$-tagged sample, excluding
events selected by the cuts defined in Section~\ref{sec:tch_cutbased}. Since the 
2-jet $b$-tagged sample includes some $t$-channel signal events, despite requiring
that the events fail the selection of the cut-based analysis, 
an uncertainty of 100\% on the expected $t$-channel single top-quark rate is assumed in this 
normalisation procedure. 
Both estimates of the $W+$jets backgrounds, the one of the NN analysis and the one
of the cut-based analysis, are in very good agreement with each other. 

The predicted and observed event yields, after the application of the selections described in Section~\ref{sec:objsel},
are given in Table~\ref{tab:evtyield}, separately for the electron and muon channels, in the 2-jet and 3-jet 
$b$-tagged samples.
\begin{table*}[!thb]
  \begin{center}
  \caption{Predicted and observed event yields, after selection, in the electron and muon 2-jet and
  3-jet $b$-tagged samples.  
  The multijet event yields are determined with data-driven technique.
  Contributions from $W$+jets events are normalised to observed data in 
  control regions as used in the cut-based analysis.
  The uncertainties on the multijet and the $W$+jets yields are also estimated 
  from data (see Section~\ref{sec:systematics}).
  All other backgrounds and the $t$-channel signal expectation are normalised to theoretical  
  cross sections. Uncertainties on these predictions are only reflecting the uncertainties
  on the theoretical cross section prediction and 
  do not include experimental uncertainties (such as the jet energy scale uncertainty, etc.).
\label{tab:evtyield}}
  \vspace*{2mm}
    \begin{tabular}
      {l@{\extracolsep{\fill}}@{\hspace{0.4cm}}r@{\extracolsep{0pt}$\:\pm\:$}l@{\hspace{0.4cm}}  
        @{\extracolsep{\fill}}r@{\extracolsep{0pt}$\,\pm\,$}l@{\hspace{0.2cm}} c|  
        @{\hspace{0.2cm}}@{\extracolsep{\fill}}r@{\extracolsep{0pt}$\,\pm\,$}l@{\hspace{0.4cm}}  
        @{\extracolsep{\fill}}r@{\extracolsep{0pt}$\,\pm\,$}l@{}}

    \hline \hline
    &\multicolumn{4}{c}{Electron}& & \multicolumn{4}{c}{Muon} \\
    &\multicolumn{2}{c}{2-jet} &\multicolumn{2}{c}{3-jet} &  
    &\multicolumn{2}{c}{2-jet} &\multicolumn{2}{c}{3-jet} \\
    \hline \hline
 single-top $t$-channel&447 & 11    &  297 & 7  & & 492 & 12  & 323 & 8 \\   
    \hline 
    $t\bar{t}$, other top &
                         785 & 52   & 1700 & 120 & & 801 &  53 &1740 &130 \\   
    $W$+light jets   &  350 &100    &  128 &  56 & & 510 & 150 & 209 & 91 \\   
    $W$+heavy flavour jets &
                       2600 &740    & 1100 & 400 & &3130 & 880 &1270 &480 \\    
    $Z$+jets, diboson & 158 & 63    &   96 &  44 & & 166 &  61 &  80 & 31 \\ 
    Multijet        &  710 &350    &  580 & 290 & & 440 & 220 & 270 &140 \\ 
      \hline	     
    Total expected        & 5050 &830    & 3900 & 520 & &5530 & 930 &3900 &520 \\ 
    \hline \hline
    Data             & \multicolumn{2}{c}{5021} & \multicolumn{2}{c}{3592} & &
                     \multicolumn{2}{c}{5592} & \multicolumn{2}{c}{3915}\\  
    \hline \hline
  \end{tabular}
 \end{center}
\end{table*}
For the purpose of this table and the histograms of kinematic distributions
the contributions of the $t\bar{t}$, $Wt$ and $s$-channel processes
have been grouped together into one category called 
``$t\bar{t}$, other top''.
In the subsequent analysis the electron and muon channels are combined.

\section{Signal and background discrimination}

To separate $t$-channel single top-quark signal events from background 
several kinematic variables are combined into one discriminant 
by employing a NN, that also exploits correlations between the variables. 
The result of the NN analysis is corroborated by a cut-based analysis
that applies additional criteria to the basic selection
described in Section~\ref{sec:objsel}.

\paragraph{Neural network based discriminant}
\label{sec:tch_NN}
The NeuroBayes~\cite{feindt-2004,Feindt:2006pm} tool (version 3.3) is used
for preprocessing the input variables and for the training of the NN. 
A large number of input variables is studied, but
only the highest-ranking variables are chosen for the training of
the NN. 
The ranking of variables is automatically determined as part of the
preprocessing step and is independent of the training procedure.
The total correlation $\kappa_\mathrm{t}^\mathrm{total}$ of a set of variables 
to the target function, that assumes
the value 1 for signal and 0 for background events, is computed as a
measure of the discrimination power of these variables. 
In an iterative procedure, the variables are sorted according to the loss
in $\kappa_\mathrm{t}^\mathrm{total}$ that is induced due to their removal from the set.
Considering the number of simulated events used to determine the
ranking, one can compute the significance of the information loss
caused by the removal of a certain variable. For the training of the NN
we use only variables that contribute with more than 
20 (10) standard deviations to $\kappa_\mathrm{t}^\mathrm{total}$
in the 2-jet (3-jet) data set.
This choice is a compromise between the achievable discrimination 
power, that increases with the number of variables, 
and the practical aim of keeping the number of variables at a manageable level.

As a result of this optimization procedure 12 kinematic variables are identified
that serve as inputs to the NN in the 2-jet data set.
The most discriminating variable is the invariant mass of the $b$-tagged jet, 
the charged lepton, and the neutrino, $m(\ell\nu b)$, which is an
estimator for the top-quark mass for signal events.
In this calculation the transverse momentum of the neutrino is given by
the $x$- and $y$-components of the $\MET$~vector, while the unmeasured 
$z$-component of the neutrino momentum, $p_z(\nu)$, is inferred by 
imposing a $W$-boson mass constraint on the lepton-neutrino system. 
Since the constraint leads to a quadratic equation for $p_z(\nu)$, 
a two-fold ambiguity arises. In the case of two real solutions, 
the one with the smaller $|p_z|$ is chosen. If the solutions are complex, 
those are avoided by a kinematic
fit that rescales the neutrino $p_x$ and $p_y$ such that the imaginary radical
vanishes, but keeps the transverse components of the neutrino as close
as possible to the $\MET$.
The second and third most discriminating variables are
the absolute value of the pseudorapidity of the highest 
$p_\mathrm{T}$ untagged jet $|\eta(j_\mathrm{u})|$
and the transverse energy of the untagged jet $E_{\mathrm{T}}\left(j_\mathrm{u}\right)$. 
Other variables used by the NN in the 2-jet data set are:
the absolute value of $\Delta \eta$ between the $b$-tagged jet
and the reconstructed $W$ boson $|\Delta\eta(b, W)|$;
the absolute value of $\Delta \eta$ between the $b$-tagged jet
and the highest $\pT$ untagged jet $|\Delta\eta(b,j_\mathrm{u})|$;
the transverse momentum of the charged lepton $p_\mathrm{T}(\ell)$; 
the scalar sum of the transverse momenta of the lepton, jets, and \MET,
$H_{\mathrm{T}}(\ell,\mathrm{jets},\MET)$; $m_{\mathrm{T}}(W)$; 
the pseudorapidity of the lepton $\eta(\ell)$; 
the invariant mass of the $b$-tagged jet $m(b)$; \MET; and the
invariant mass of the untagged jet and the $b$-tagged jet
$m(j_\mathrm{u}b)$.

\begin{figure*}[!h!tpb]
 \centering
 \subfigure[]{
   \includegraphics[width=0.30\textwidth]{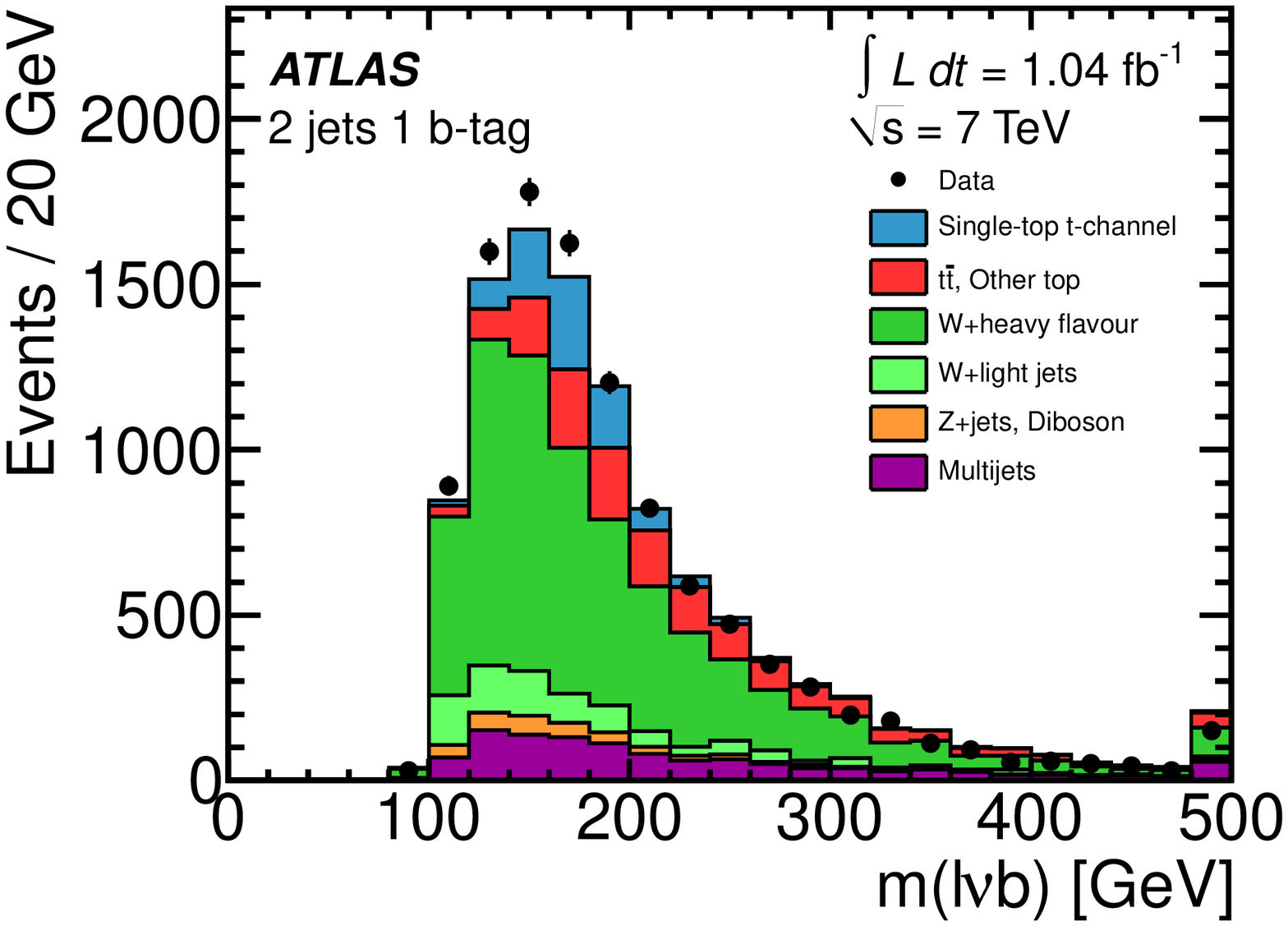}
 } 
 \subfigure[]{
   \includegraphics[width=0.30\textwidth]{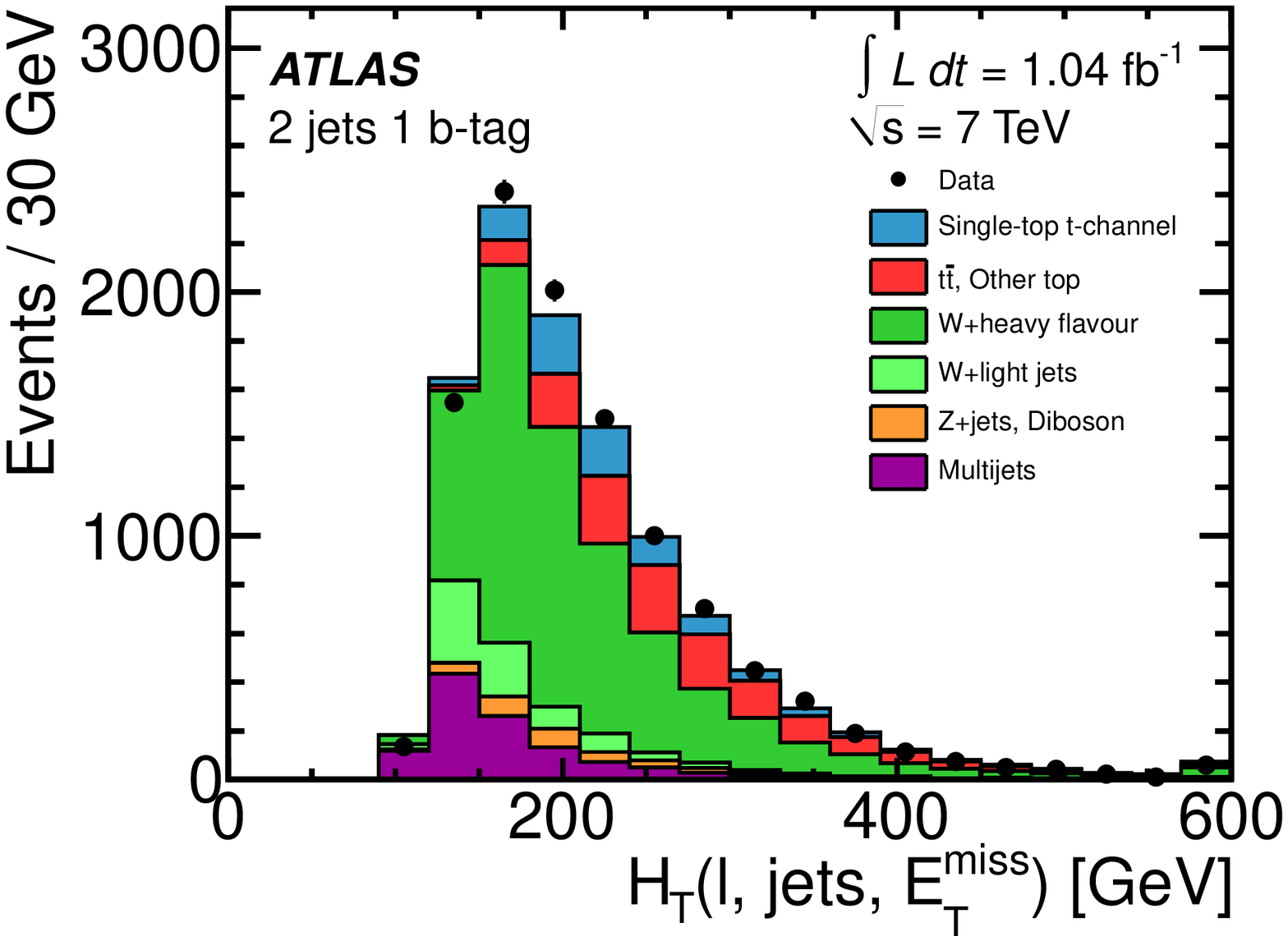}
 }

 \vspace*{2mm}

 \subfigure[]{
   \includegraphics[width=0.30\textwidth]{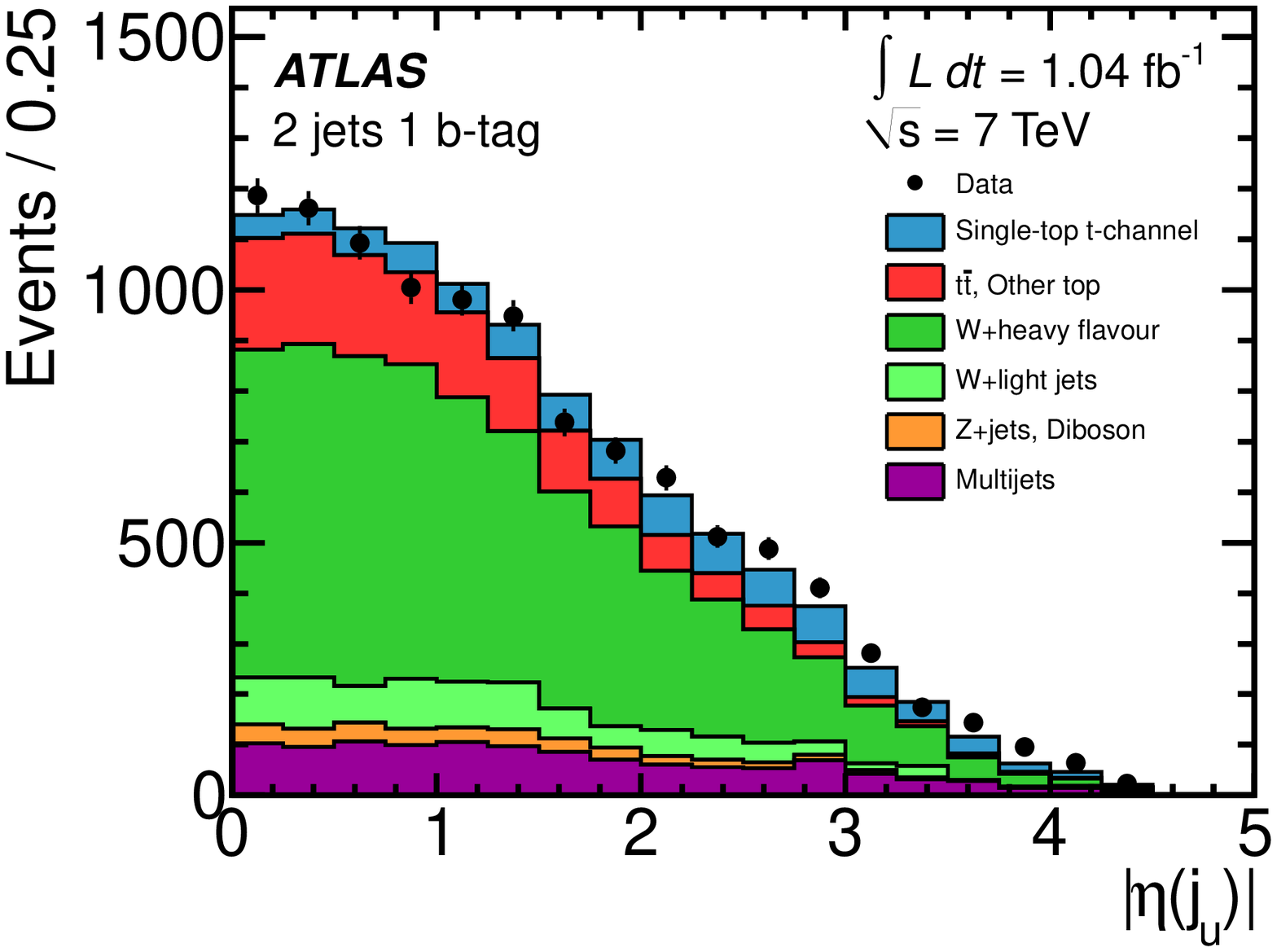}
 }
 \subfigure[]{
   \includegraphics[width=0.30\textwidth]{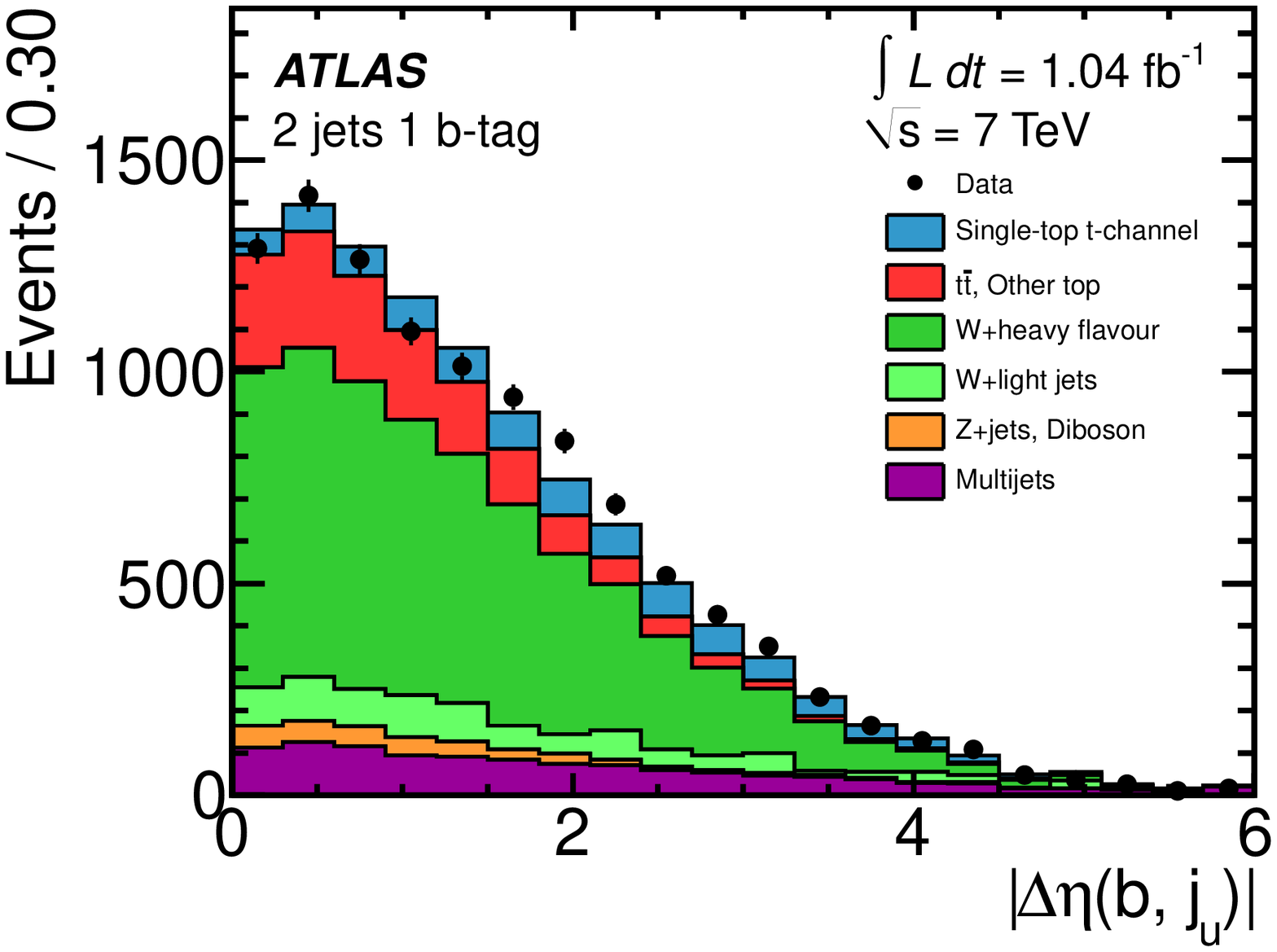}   
 }

 \vspace*{2mm}

 \subfigure[]{
  \includegraphics[width=0.30\textwidth]{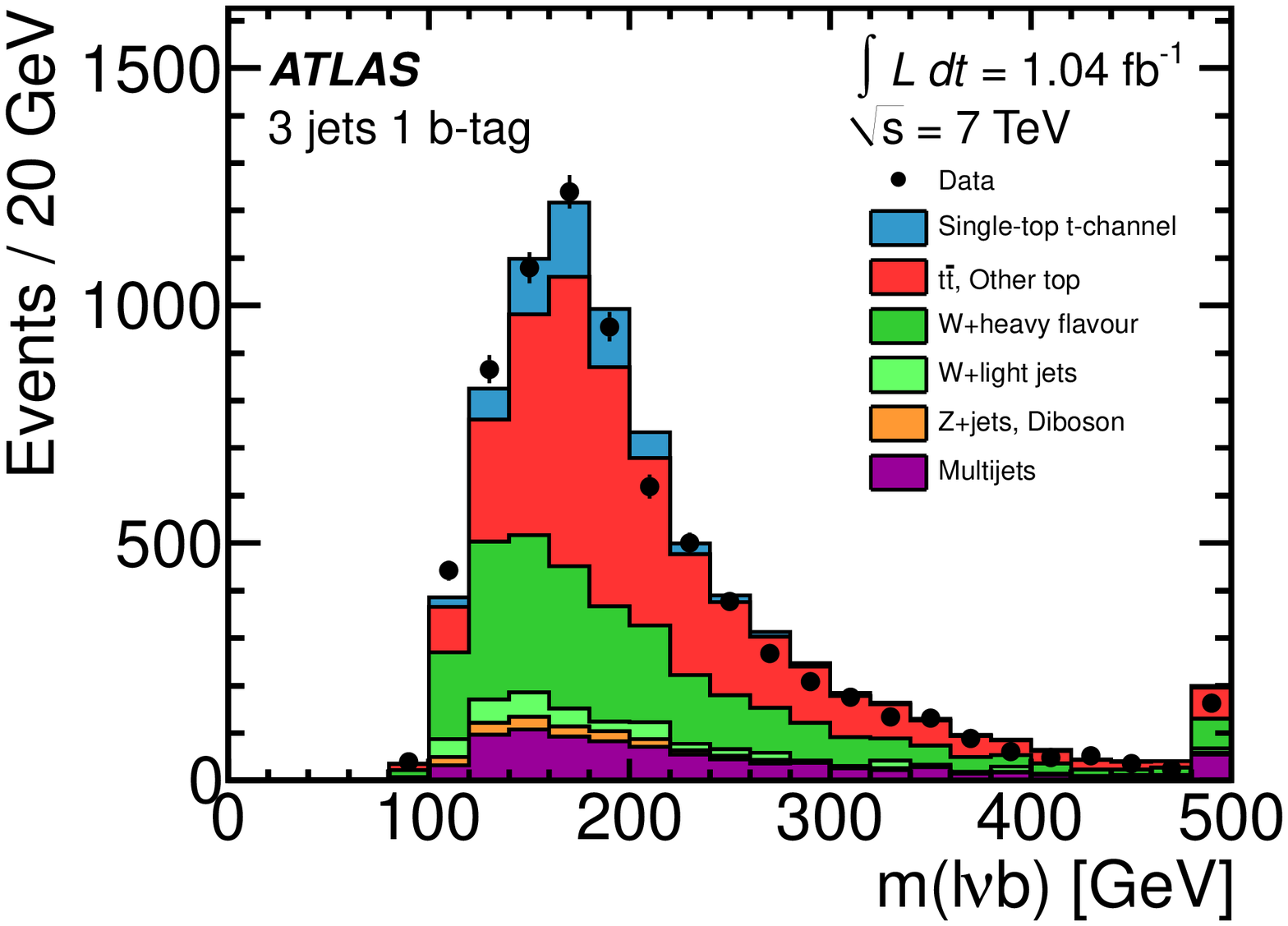}
 }
 \subfigure[]{
  \includegraphics[width=0.30\textwidth]{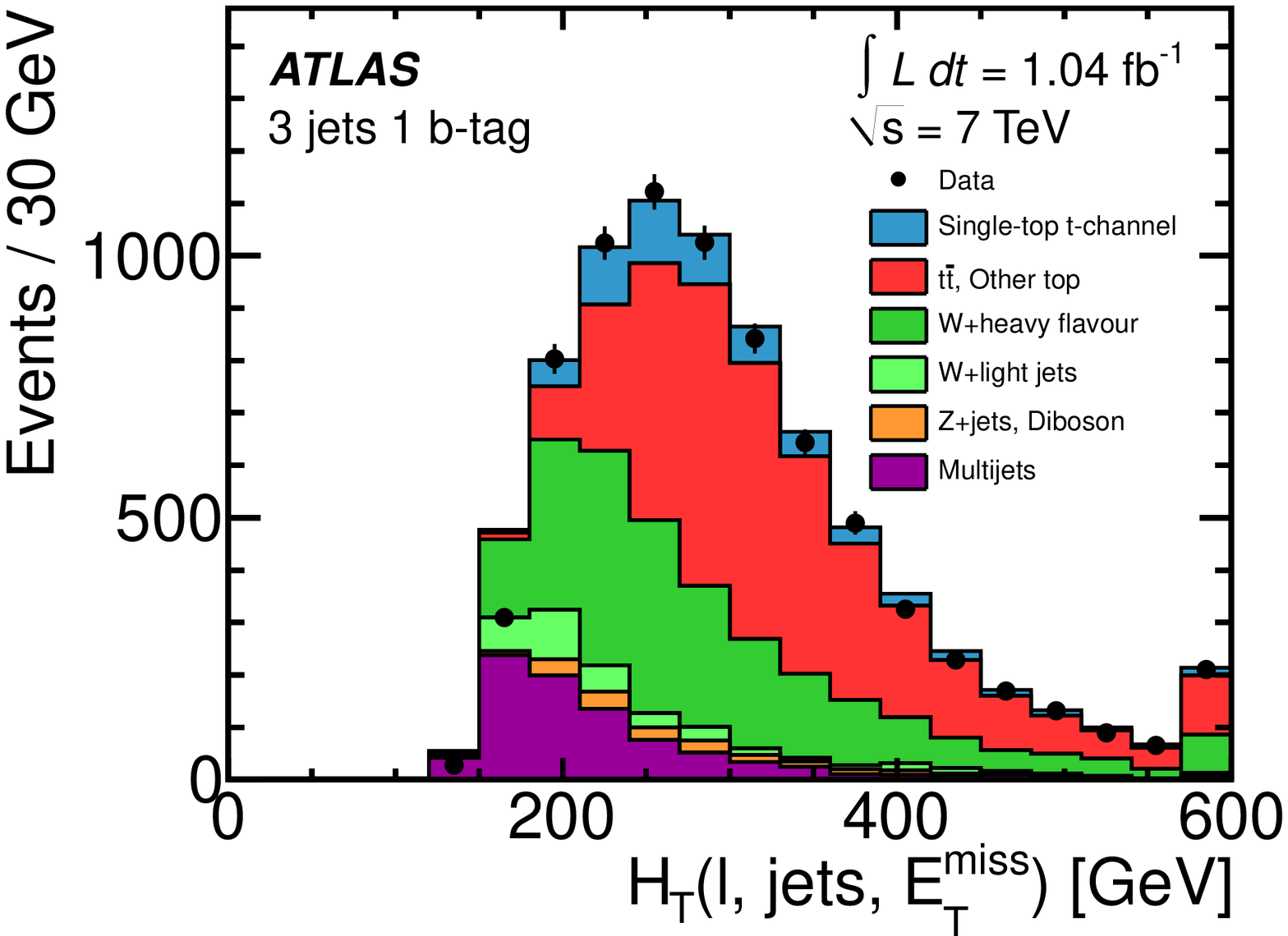}
 }

 \vspace*{2mm}

 \subfigure[]{
   \includegraphics[width=0.30\textwidth]{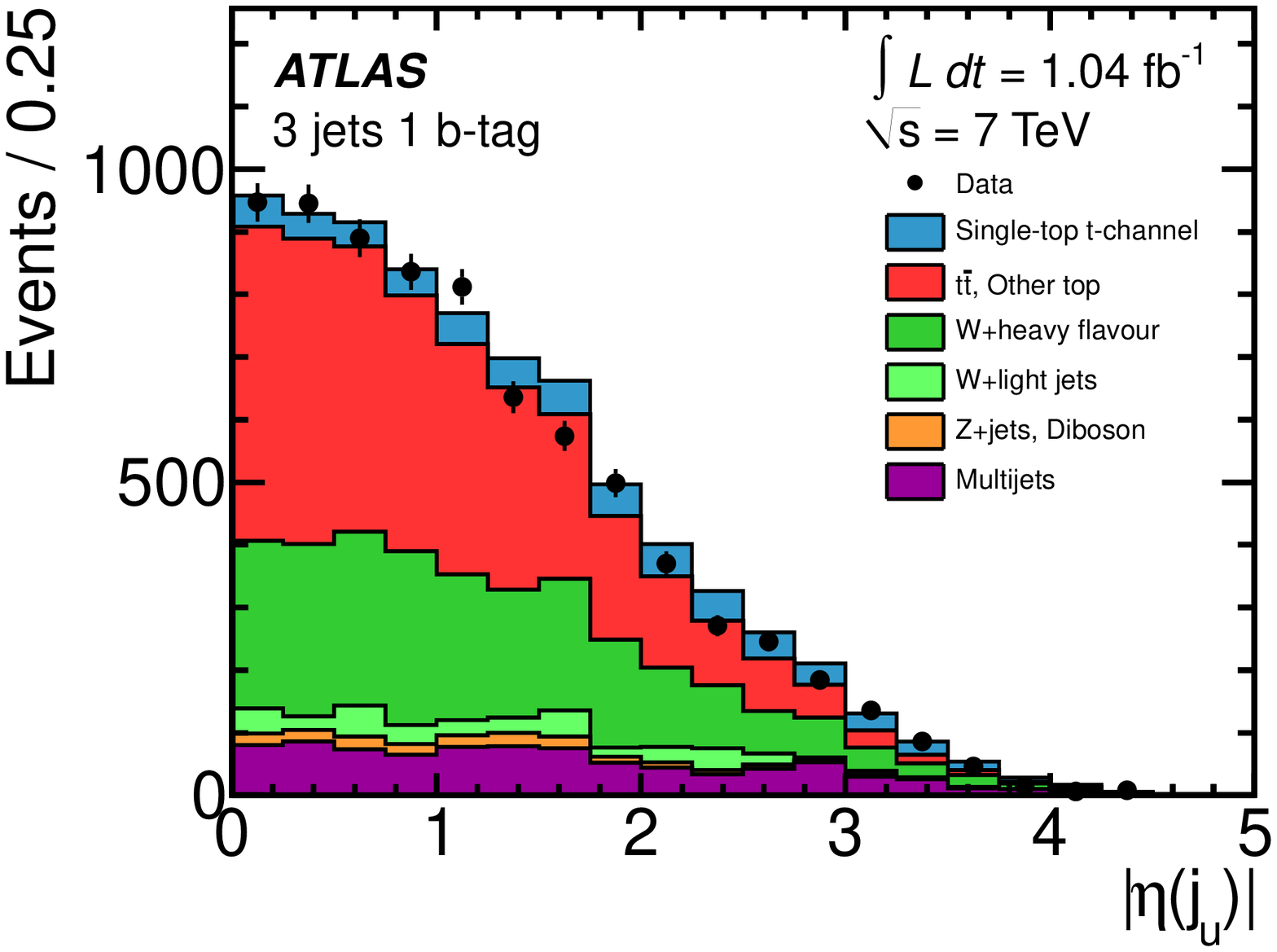}
 }
 \subfigure[]{
   \includegraphics[width=0.30\textwidth]{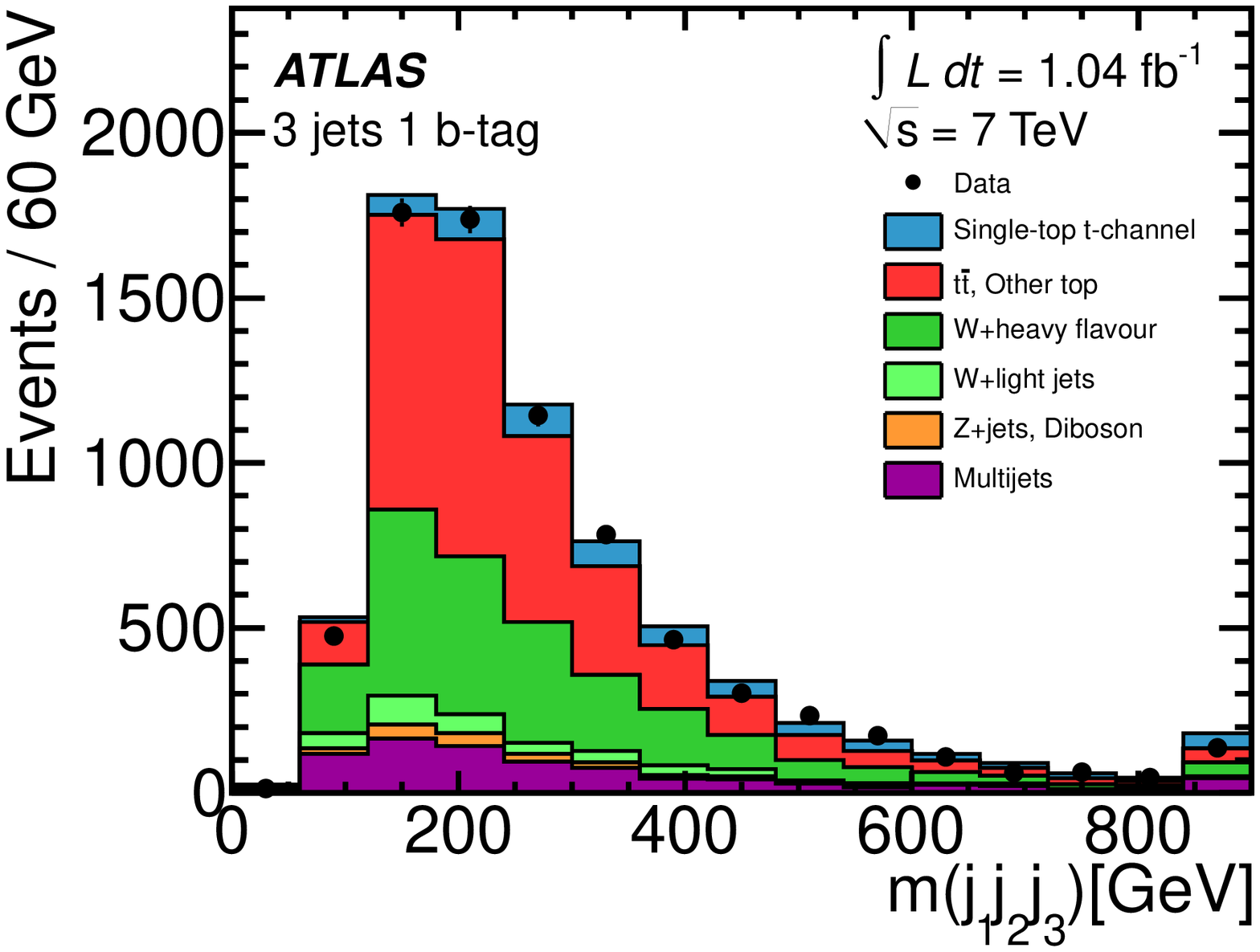}
 }
\caption{Discriminating variables in the $b$-tagged sample for 2-jet events and 3-jet events.
  Multijet event yields are determined with data-driven techniques.
  Contributions from $W$+jets events are derived from simulation and
  normalised to data in control regions with the method employed in the cut-based analysis.
  All other backgrounds and the $t$-channel signal expectation are normalised to theoretical  
  cross sections.
  The last histogram bin includes overflows.
  The figures shown are for 2-jet or 3-jet events, respectively:
  (a), (e) the invariant mass of the $b$-tagged jet, the charged lepton, and the neutrino;
  (b), (f) the scalar sum of the transverse momenta of the lepton, the jets, and \MET;
  (c), (g) the absolute value of the pseudorapidity of the highest $p_\mathrm{T}$ untagged jet.
  For 2-jet events Figure (d) shows the absolute value of $\Delta \eta$ between the $b$-tagged jet 
  and the highest $\pT$ untagged jet; and for 3-jet events Figure
  (h) displays the invariant mass of the three selected jets.
}
 \label{fig:Plot_CB_DataMC}
 \end{figure*}

\begin{figure*}[!t]
\begin{center}
  \subfigure[]{
    \includegraphics[width=0.35\textwidth]{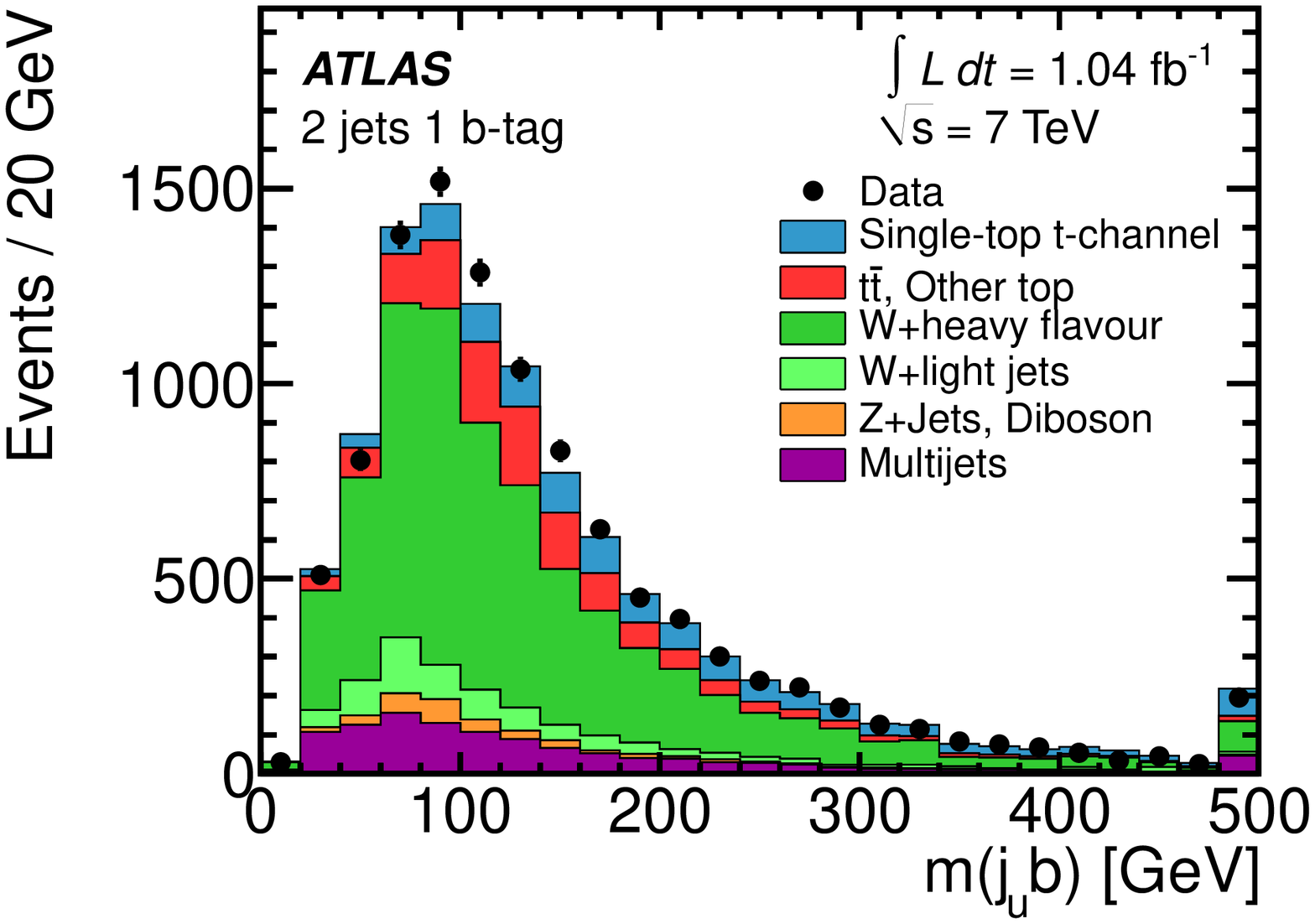}
  }
  \subfigure[]{
    \includegraphics[width=0.35\textwidth]{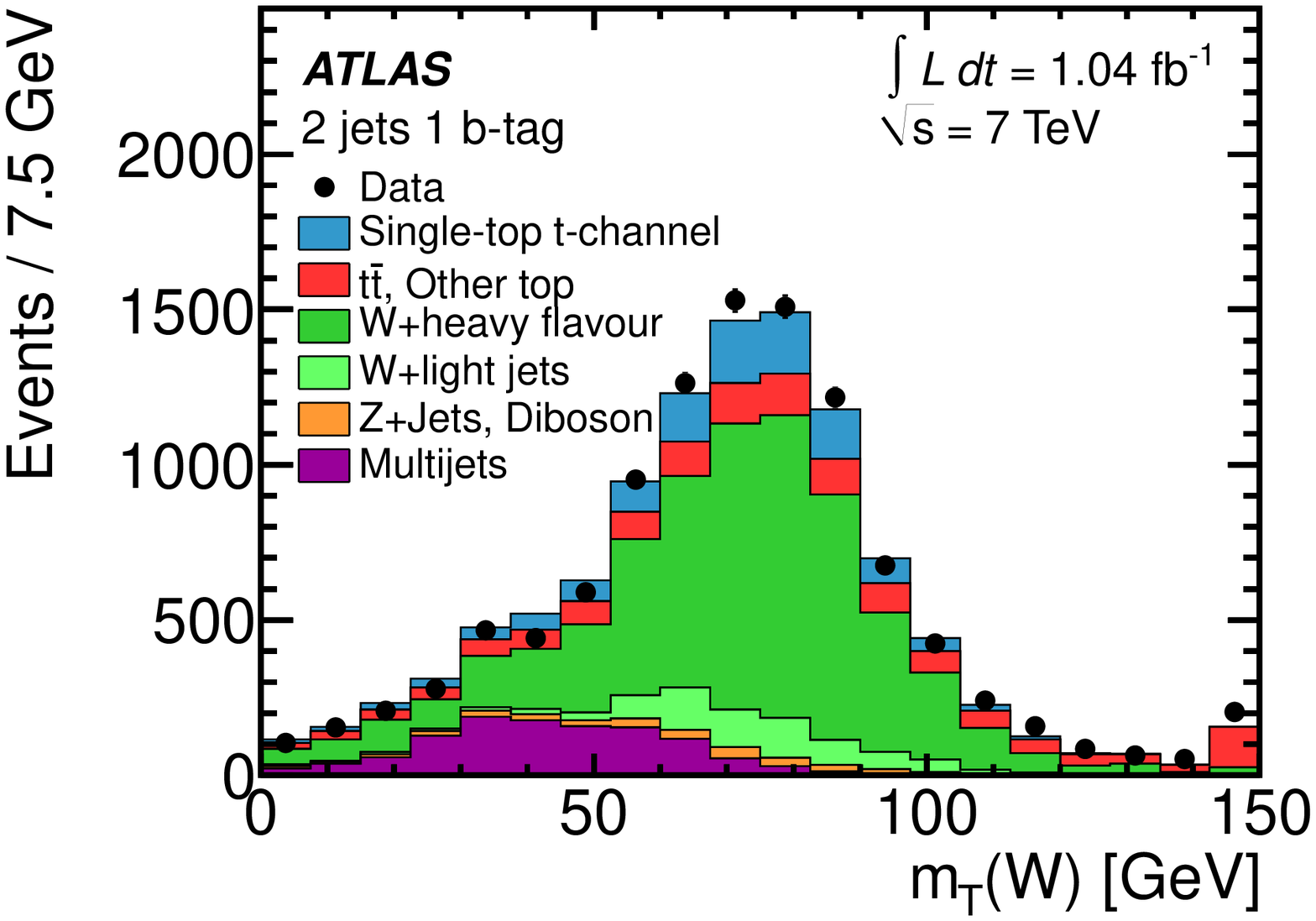}
  }
  \vspace*{2mm}
  \subfigure[]{
    \includegraphics[width=0.35\textwidth]{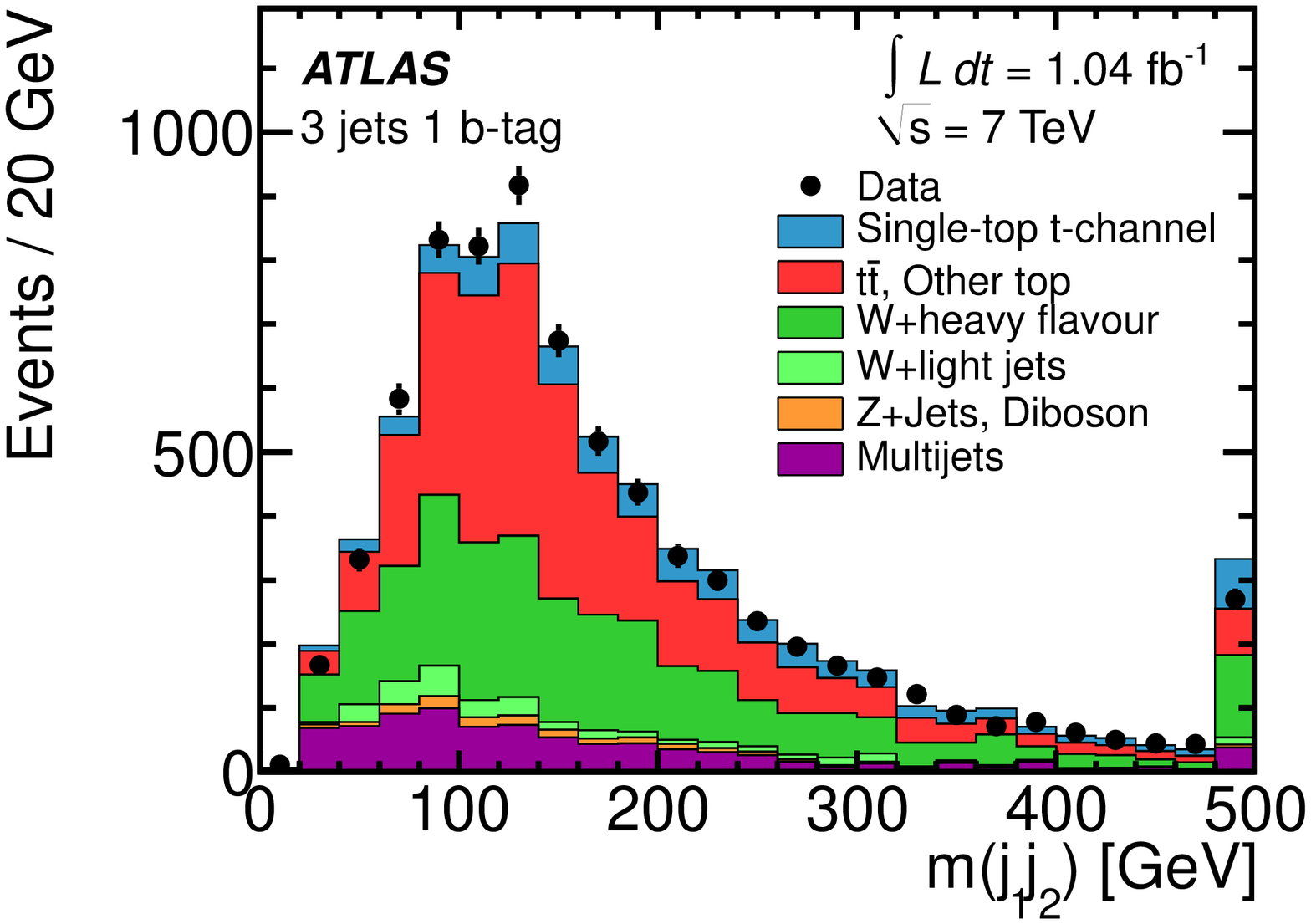}
  }
  \subfigure[]{
    \includegraphics[width=0.35\textwidth]{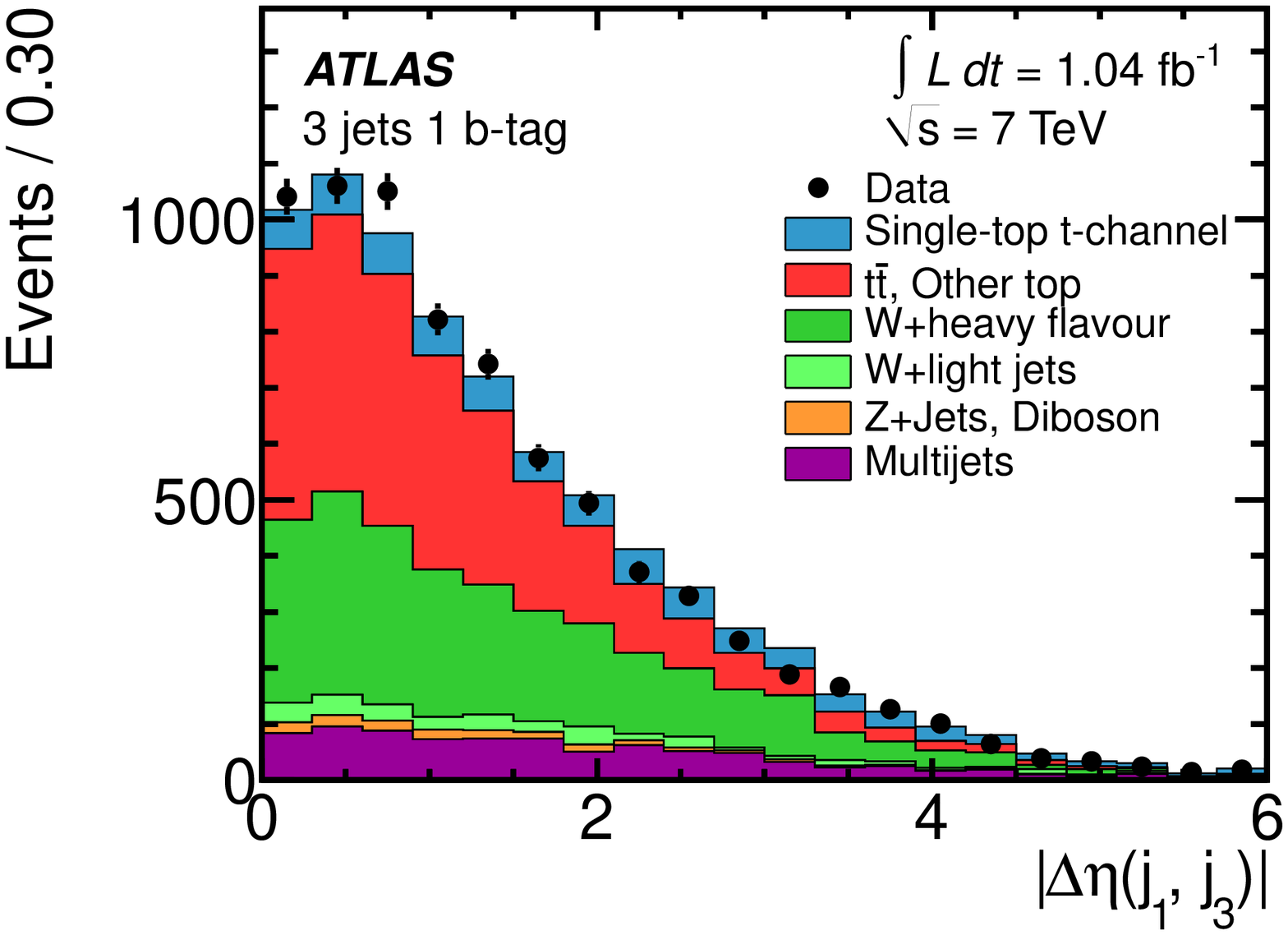}
  }
\caption{\label{fig:NN_2j_3jtag}
Additional discriminating variables used in the NN analysis for 2-jet and 3-jet events.
The rate of multijet events is normalized to the estimate obtained from the fit
to the $\MET$ distributions.
All other component distributions are normalised to the result of the maximum-likelihood 
fit of the NN output.
The last histogram bin includes overflows.
The figures shown are, for 2-jet events:
(a) invariant mass of the highest $p_\mathrm{T}$ untagged jet and the $b$-tagged jet;
(b) transverse mass of the lepton-\MET system;
and for 3-jet events:
(c) invariant mass of the two leading jets;
(d) absolute value of $\Delta \eta$ between the leading jet and the lowest $\pT$ jet.
}
\end{center}
\end{figure*}

For events with three jets 18 variables are used, the most
discriminating ones being the invariant mass of the two leading jets, $m(j_1j_2)$,
$m(\ell\nu b)$, and the absolute value of the difference in the pseudorapidity of the leading and lowest $p_\mathrm{T}$ jet,
$|\Delta\eta(j_1,j_3)|$.
Figure~\ref{fig:Plot_CB_DataMC} shows distributions of some of the most 
discriminating variables in the $b$-tagged 2-jet or 3-jet samples, 
used in both the NN analysis and the cut-based analysis.
The variable $m(j_1j_2j_3)$ denotes the invariant mass of all selected jets
in the 3-jet data set. 
Distributions of additional variables used only in the NN approach 
are shown in Figure~\ref{fig:NN_2j_3jtag}.

\begin{figure*}[!t]
\begin{center}
  \subfigure[]{
    \includegraphics[width=0.35\textwidth]{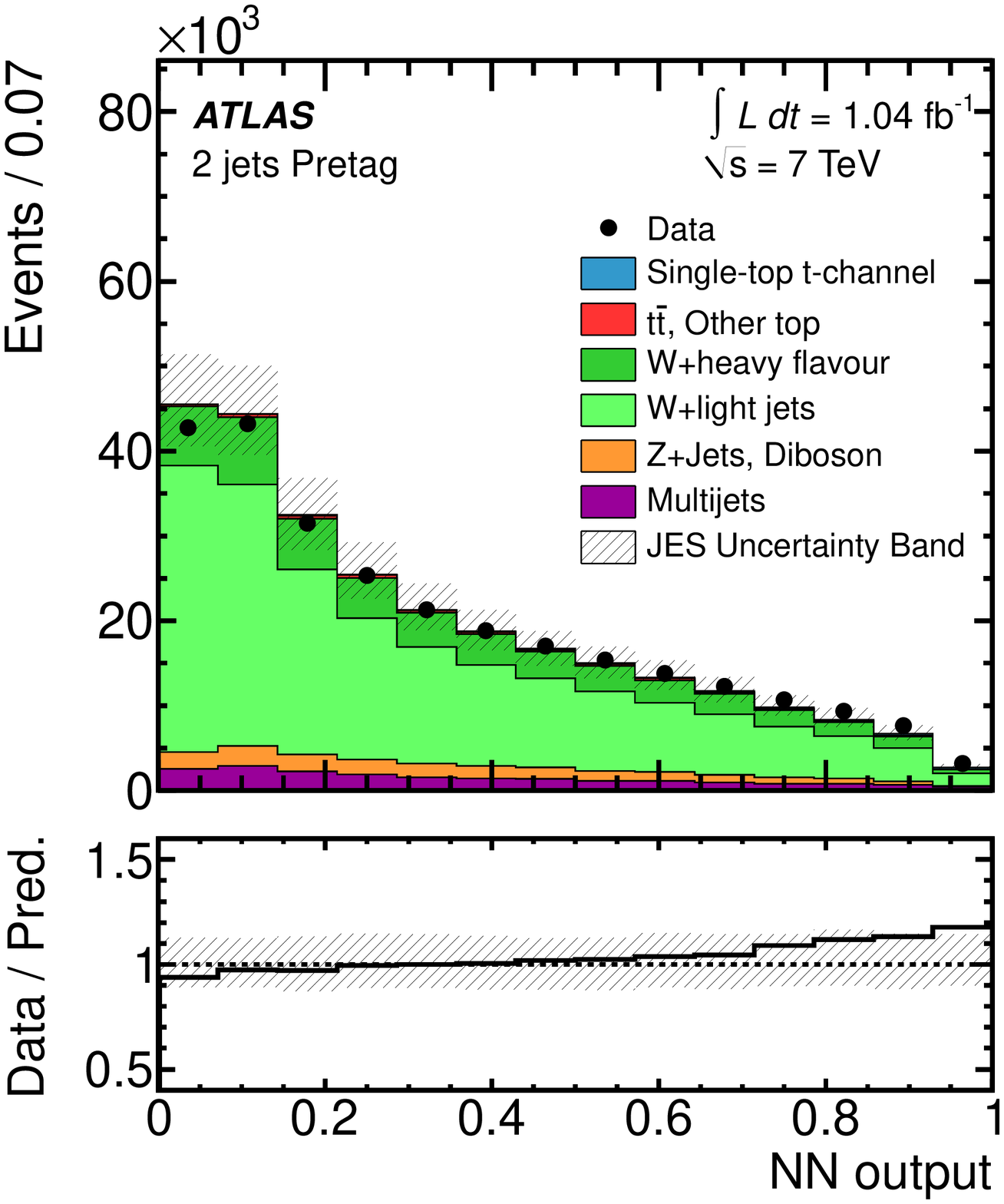}
    \label{fig:NNout_2j0tag}
  }
  \subfigure[]{
    \includegraphics[width=0.35\textwidth]{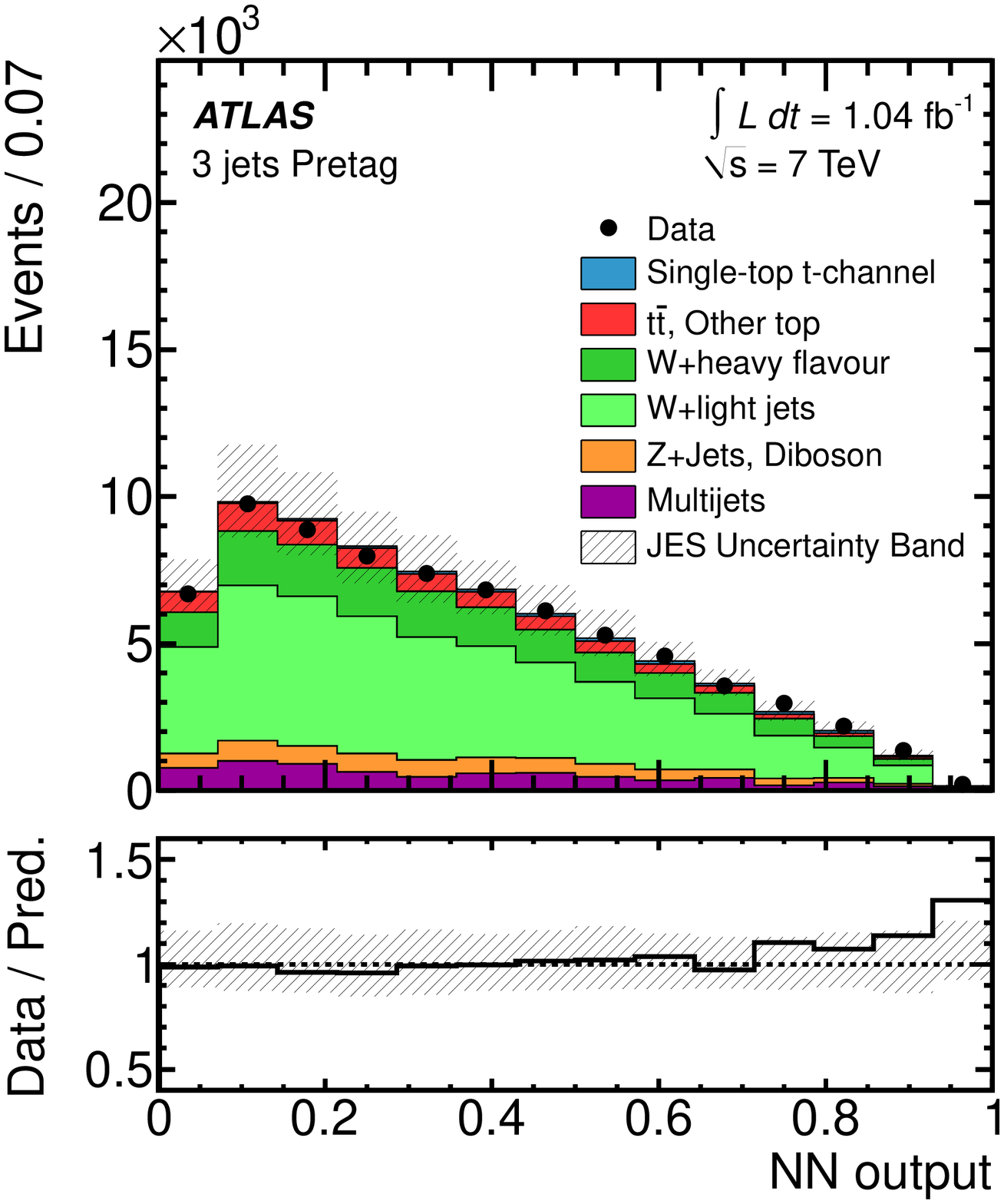}
    \label{fig:NNout_3j0tag}
  }
  \vspace*{1mm}

  \subfigure[]{
    \includegraphics[width=0.35\textwidth]{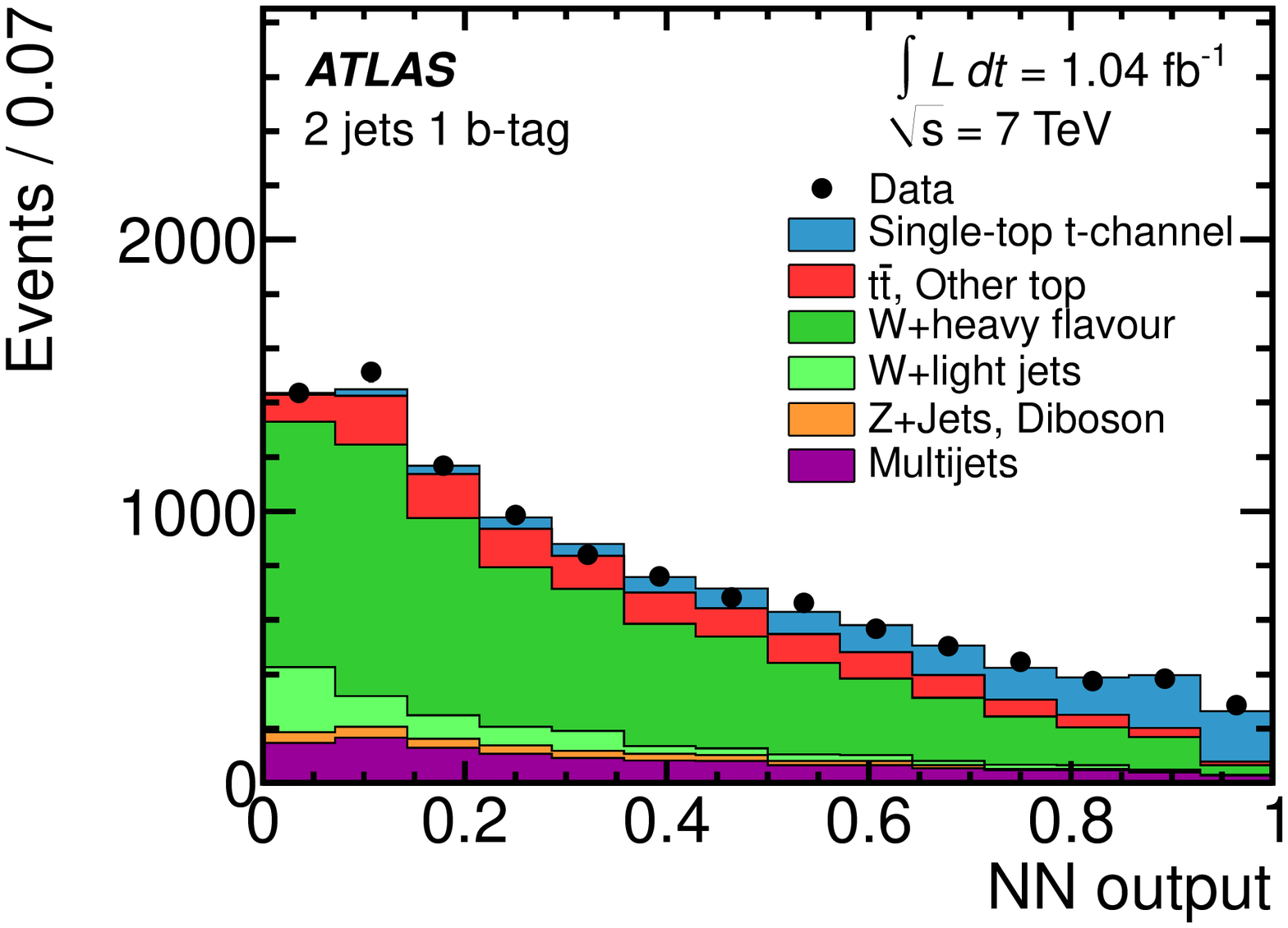}
    \label{fig:NNout_2j1tag}
  }
  \subfigure[]{
    \includegraphics[width=0.35\textwidth]{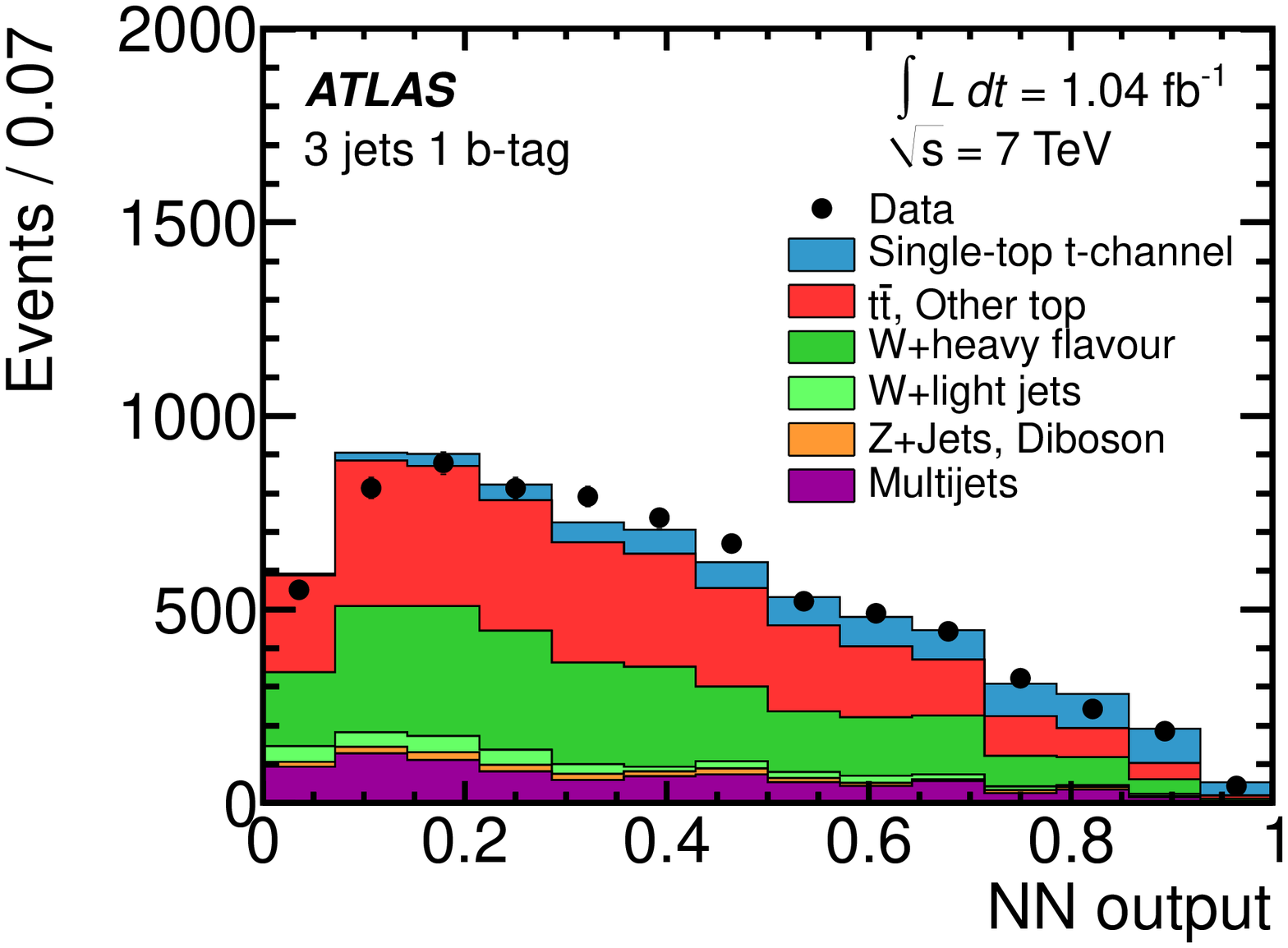}
    \label{fig:NNout_3j1tag}
  }
  \caption{\label{fig:NNoutput}
  (a) and (b): Neural network output distribution 
  for the pretag sample, including the JES uncertainty on the prediction (hatched region).
  The multijet component is normalized to the estimate obtained 
  from the fit to the $\MET$ distributions.
  All other components are normalised such that the total number of expected events 
  in the pretag sample is equal to the observed number of events.
  The ratio beween the data and the total predicted distributions is also shown.
  (c) and (d): NN output distribution for the 2-jet and 3-jet $b$-tagged samples, respectively.
  All component distributions are normalised to the result of the maximum-likelihood fit,
  except for the component of multijet events that is normalized to the estimate obtained 
  from the fit to the $\MET$ distributions.
}
\end{center}
\end{figure*}

The agreement between the background model and collision data is 
tested in the large pretag sample for each input variable used in the analysis, 
for various additional control variables, and the NN output distributions, 
which are shown in Figs.~\ref{fig:NNout_2j0tag} and \subref{fig:NNout_3j0tag}.
In this control sample, where the $b$-tagging algorithm has not yet been applied,
the $b$-tagged jet is substituted by the most central jet, with the requirement 
that it is within $|\eta|<2.5$. 
Good agreement is found overall, except for the $|\eta|$ distribution of the
jet with the highest $|\eta|$ in the pretag
data set for which an additional systematic modelling uncertainty is taken 
into account (see Section~\ref{sec:systematics}).

The NeuroBayes tool
combines a three-layer feed-forward NN with a complex preprocessing of the input
variables. By transforming the variables in the preprocessing step
the influence of outliers is largely reduced and statistical fluctuations are damped.
NeuroBayes applies Bayesian regularisation techniques for the 
training process to damp statistical fluctuations in the training sample
and to avoid overtraining.
A certain fraction of simulated events (20\%) is not included in the training
sample and is used as an independent test sample to check that there is no
overtraining.
The ratio of signal to background events in the training is 
chosen to be 1:1, while the different background processes are weighted according 
to the number of expected events.

To extract the signal content of the selected sample 
a maximum-likelihood fit is performed to the complete NN output distributions 
in the 2-jet and 3-jet data sets (see Section~\ref{sec:xs_measure}).
Fitting all bins of the distribution has the advantage of 
making maximal use of the signal events remaining after the 
event selection, and also allows the background rates to be
constrained by the data. The sensitivity to the background
rates is given by the background dominated region close to zero.
The observed NN output distributions scaled to the fit result are shown in 
Figs.~\ref{fig:NNoutput}\subref{fig:NNout_2j1tag} and \ref{fig:NNoutput}\subref{fig:NNout_3j1tag}
for $b$-tagged events with two or three jets, respectively.

\paragraph{Cut-based selection}
\label{sec:tch_cutbased}
In the cut-based analysis additional selections are applied to a
subset of five variables used by the NN analysis:
$|\eta(j_\mathrm{u})|>2$, $H_{\mathrm{T}}(\ell,\mathrm{jets},\MET)>210\gev$, 
and $150\gev<m(\ell\nu b)<190\gev$.  
The 2-jet selection requires $|\Delta\eta(b,j_\mathrm{u})| > 1$, 
while the 3-jet selection requires that $m(j_1j_2j_3)$ is higher than $450\gev$, 
to further reduce the large \ttbar~ contribution in this channel.  
The selection cuts were chosen 
in order to increase the expected significance of the $t$-channel single 
top-quark signal, taking into account systematic uncertainties 
on the background estimate~\cite{GaussSideband}.

The positive and negative lepton-charge samples are considered
separately, as more single-top quark $t$-channel events are expected 
in the $e^{+}/\mu^{+}$ samples than in the $e^{-}/\mu^{-}$ 
samples due to the dominance of valence $u$~quarks in the proton over $d$~quarks. 
The 2-jet and 3-jet data sets are also considered separately.

\begin{figure*}[hbtp]
 \centering
  \subfigure[]{
    \includegraphics[width=0.30\textwidth]{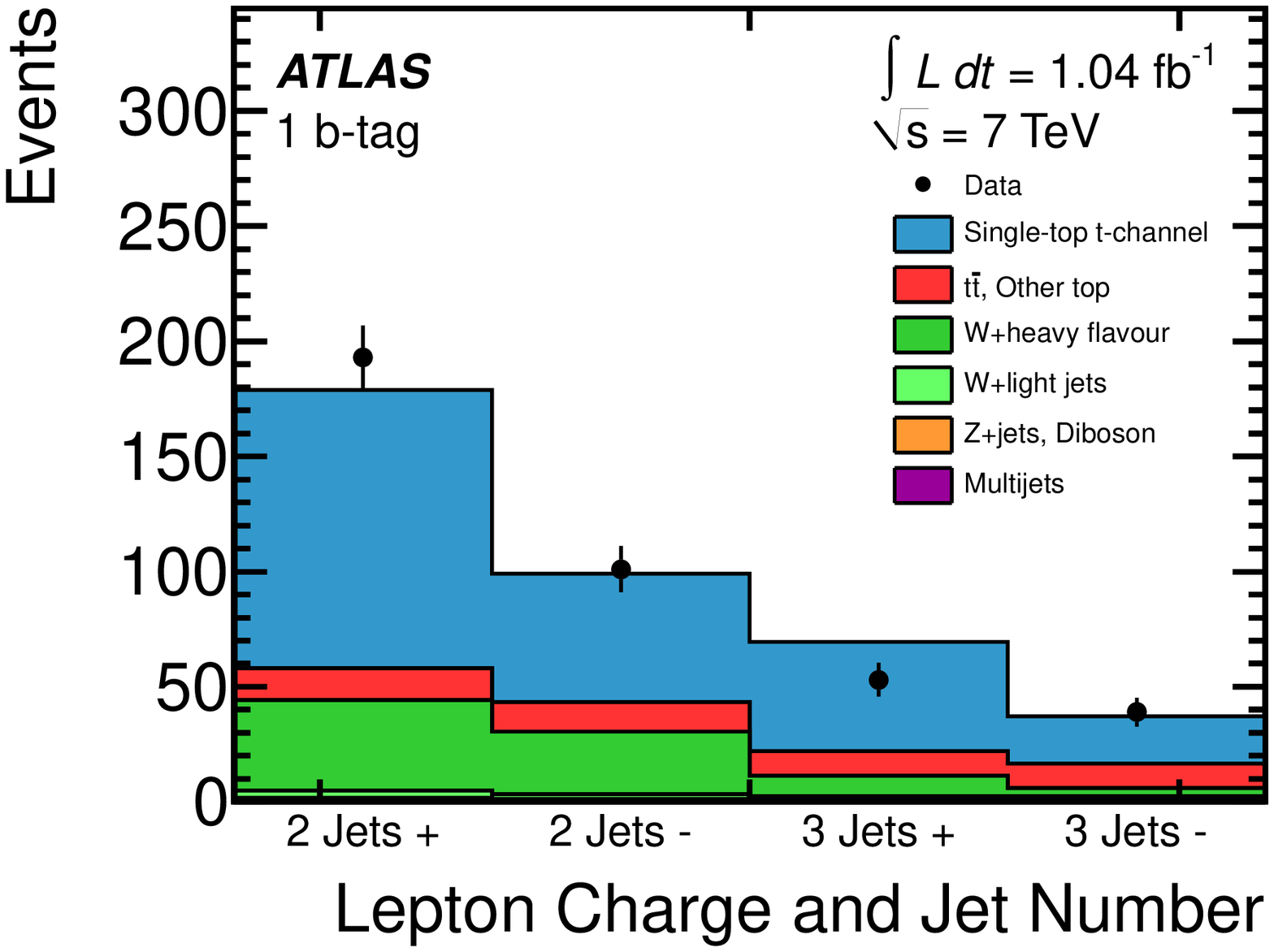}
    \label{fig:CB_leptcharge_jet_multiplicity}
  }
  \subfigure[]{
    \includegraphics[width=0.30\textwidth]{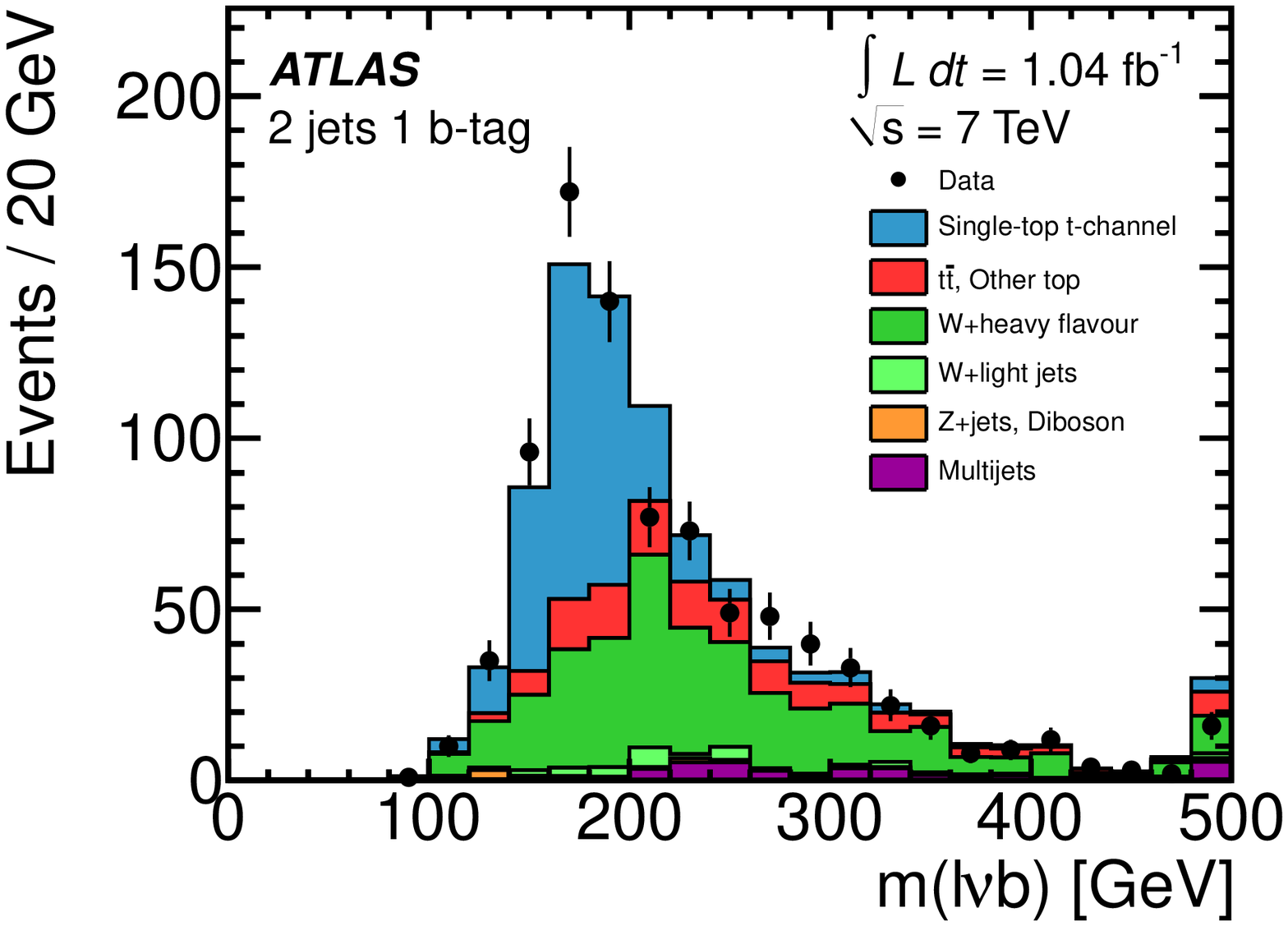}
    \label{fig:CB_mlnub_aftercuts_2j} 
  }
  \subfigure[]{
    \includegraphics[width=0.30\textwidth]{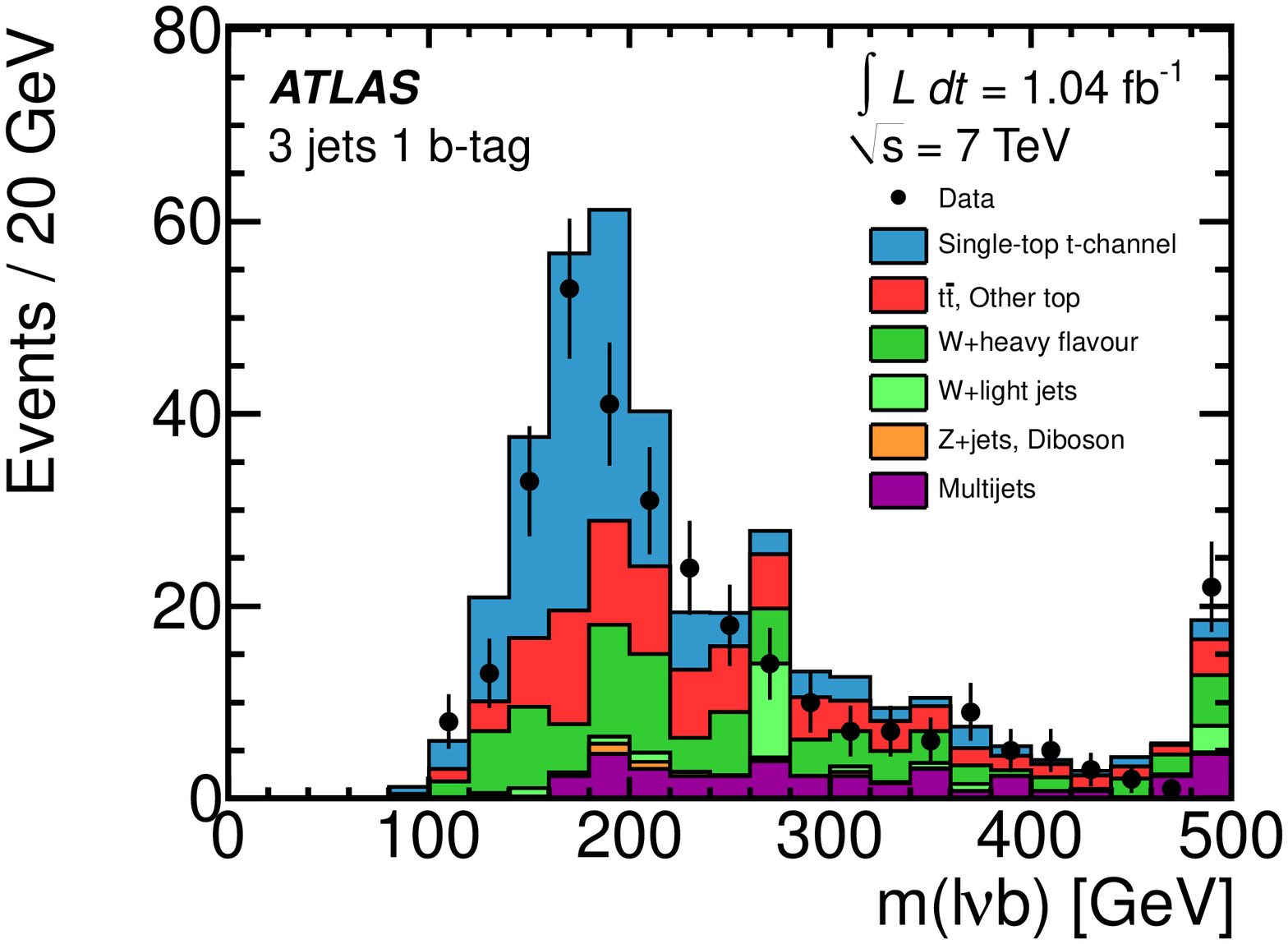}
    \label{fig:CB_mlnub_aftercuts_3j}
}
 \caption{
 (a) Distribution of the lepton charge after the full cut-based selection for 2-jet and 3-jet events.
 (b,c) invariant mass of the $b$-tagged jet, the charged lepton, and the neutrino, $m(\ell\nu b)$, for the $b$-tagged sample for 2-jet (b) and 3-jet (c) events 
 after applying all cut-based selections except for the cut on $m(\ell\nu b)$. 
 In all three distributions the $t$-channel single top-quark contribution is normalised to the observed cross section
 obtained with the cut-based analysis.
 The last histogram bin includes overflows.
}
 \label{fig:tch_results}
 \end{figure*}

The signal and background event counts for the positive and negative lepton-charge samples 
are given in Table~\ref{tab:tch_eventyields_CB}.
The observed event yields are consistent with the SM expectation in each channel.
Figure~\ref{fig:tch_results}\subref{fig:CB_leptcharge_jet_multiplicity} 
shows the distribution of the lepton charge for events with two or three jets
after the application of all cut-based selections.
Figures~\ref{fig:tch_results}\subref{fig:CB_mlnub_aftercuts_2j} 
and~\ref{fig:tch_results}\subref{fig:CB_mlnub_aftercuts_3j} display the
variable $m(\ell\nu b)$ for 2-jet and 3-jet events respectively
after applying all selections except for the cut on $m(\ell\nu b)$.
In these figures, the $t$-channel single top-quark contribution is normalised to the observed cross section 
as measured from the combination of all four channels.
\begin{table*}[!h!tpb]
  \begin{center}
 \caption{Event yield for the 2-jet and 3-jet $b$-tagged positive and negative lepton-charge channels after the 
cut-based selection. 
  The multijet and $W$+jets backgrounds are normalised to observed data in 
  control regions, all other samples are normalised
to theory cross sections. Uncertainties shown include all sources of systematic errors,
summed quadratically and without taking into account possible anticorrelations between systematic sources and between processes.
\label{tab:tch_eventyields_CB}}
    \vspace*{2mm}
     \begin{tabular}
      {l@{\extracolsep{\fill}}@{\hspace{0.4cm}}r@{\extracolsep{0pt}$\:\pm\:$}l@{\hspace{0.4cm}}  
        @{\extracolsep{\fill}}r@{\extracolsep{0pt}$\,\pm\,$}l@{\hspace{0.2cm}} c|  
        @{\hspace{0.2cm}}@{\extracolsep{\fill}}r@{\extracolsep{0pt}$\,\pm\,$}l@{\hspace{0.4cm}}  
        @{\extracolsep{\fill}}r@{\extracolsep{0pt}$\,\pm\,$}l@{}}
    \hline \hline

    &\multicolumn{4}{c}{Cut-based 2-jet}& & \multicolumn{4}{c}{Cut-based 3-jet} \\
    &\multicolumn{2}{c}{Lepton +} &\multicolumn{2}{c}{Lepton --} &  
    &\multicolumn{2}{c}{Lepton +} &\multicolumn{2}{c}{Lepton --} \\

    \hline \hline
    single-top $t$-channel &  85 & 29  &  39 & 13  &&  33.6 & 7.0 &  14.6 & 6.2  \\
    \hline                                                                       
    $t\bar t$, other top   &  14.0 & 6.4  &  12.8 & 4.2  &&  10.5 & 4.2  &  10.7 & 7.9  \\
    $W$+light jets         &  3.3 & 1.9    &  2.0 & 1.2  &&  0.8 & 1.3  &  0.3 & 0.3  \\
    $W$+heavy flavour jets &  39 & 11   &  27.1 & 7.5  &&  8.7 & 6.0  &  3.4 & 3.1  \\
    $Z$+jets, diboson      &  1.1 & 0.8   &  1.0 & 0.8  &&  0.3 & 0.2  &  0.2 & 0.3  \\
    Multijet              &  0.2 & 0.2  &  0.3 & 0.3  &&  1.5 & 1.1  &  3.1 & 2.0  \\
    \hline    
    Total expected         &  143 & 31  &  83 & 16  &&  56 & 10  &  32 & 11  \\
    S/B                    &  \multicolumn{2}{c}{1.5}  & \multicolumn{2}{c}{0.9} &&  \multicolumn{2}{c}{1.6} & \multicolumn{2}{c}{1.0}  \\
    \hline \hline
    Data                   &   \multicolumn{2}{c}{193}  &   \multicolumn{2}{c}{101}   &&   \multicolumn{2}{c}{53}  & \multicolumn{2}{c}{39}   \\
    \hline \hline
    \end{tabular}
  \end{center}
\end{table*}

\section{Systematic uncertainties}
\label{sec:systematics}

Systematic uncertainties on the normalisation of the individual
backgrounds and on the signal acceptance affect the measured
single top-quark $t$-channel cross section. In the NN analysis the
shape of each individual prediction is also affected; both the rate and the 
shape uncertainties are taken into account by generating correlated 
pseudo-experiments. The impact of the systematic uncertainties on the
$t$-channel cross-section measurement is estimated from these pseudo-experiments.
The uncertainties can be split into the following categories:

\vspace{-0.3cm}
\paragraph{Object modelling}
Systematic uncertainties due to the residual differences between data and 
Monte-Carlo simulation for the reconstruction and energy calibration of 
jets, electrons and muons are propagated in the analysis.
The main source of object modelling uncertainty comes from the jet energy scale (JES), 
including the modelling of pile-up, as well as $b$-jet identification. 
Other components include lepton energy scale and lepton and jet identification 
efficiencies.
The JES uncertainty has been evaluated using 2010 data~\cite{ATLAS-JES-2011}. 
Additional contributions to this uncertainty 
due to the larger pile-up effects in 2011 data are included and range 
from less than 1\% to 5\% as a function of jet \pT~ and $\eta$.  
For $b$-quark jets a JES uncertainty of 0.8\% to 2.5\%, 
depending on the jet \pT, is added in quadrature to the JES uncertainty. 
Scale factors, determined from collision data~\cite{ATLAS-CONF-2011-102}, 
are applied to correct the $b$-tagging performance in simulated events to match the data.
Both $b$-jets and $c$-jets in simulation use the same $b$-tagging scale factors 
with uncertainties that depend on the \pT~ and $\eta$ of the jet. 
The uncertainties on the scale factors
vary from 10\% to 15\% for $b$-quark jets and from 20\% to 30\% for $c$-quark jets.
For light-quark jets the mis-tagging uncertainty ranges from 20\% to 50\% 
as a function of jet \pT~ and $\eta$. 
Other minor uncertainties are assigned to the reconstruction of \met\ and to account
for the impact of pile-up collisions on \met.
Finally, a systematic uncertainty was also assigned to account for temporary failures 
of parts of the LAr calorimeter readout during part of the data-taking period, 
which was not modelled in the MC samples.

\vspace{-0.3cm}
\paragraph{Monte-Carlo generators and PDFs} 
Systematic uncertainties arising from the modelling of the single top-quark signal
and the $\ttbar$ background are taken into account. 
The largest contributions come from the modelling of parton showers and hadronisation,
estimated by interchanging the modelling between \textsc{PYTHIA} and \textsc{HERWIG},
and from the amount of initial-state and final-state radiation (ISR/FSR), estimated
using dedicated \textsc{AcerMC} samples interfaced to \textsc{PYTHIA}
where parameters controlling the ISR/FSR emission are varied in a range\footnote{
The default {\textsc PYTHIA} values of these parameters are :
PARP(67) $=4.0$ and PARP(64) $=1.0$ for ISR and PARP(72) $=0.192\gev$, PARJ(82) $=1.0\gev$ for FSR.
To decrease (increase) ISR, the parameters PARP(67) and PARP(64) are set to 0.5 and 4.0 (6.0 and 0.25), respectively.
To decrease (increase) FSR, the parameters PARP(72) and PARJ(82) are set to $0.096\gev$ and $2.0\gev$ 
($0.384\gev$ and $0.5\gev$), respectively.
Samples of simulated events are produced with six different sets of parameters settings:
ISR increased (decreased), FSR increased (decreased), and a simultaneous increase (decrease) 
of ISR and FSR.
}
consistent with those used in the Perugia Hard/Soft tune
variations~\cite{PerugiaTunes}.
The uncertainty due to the choice of the
single top-quark $t$-channel signal generator is estimated from the difference between
\textsc{AcerMC} and \textsc{MCFM} predictions~\cite{Campbell:2009ss}. The modelling 
uncertainty for the $\ttbar$ background is evaluated by comparing the
generators \textsc{MC@NLO} and \textsc{POWHEG}~\cite{Alioli:2010xd,Frixione:2007nw} 
(with \textsc{HERWIG} showering).
For the $W$+jets background a shape uncertainty is assigned based on the variation of
the choices of the matching scale and of the functional form of the factorisation scale in \textsc{ALPGEN}.
Systematic uncertainties related to the parton distribution functions
are taken into account for the signal and for all background processes which are modelled
by simulated events. 
In addition to the nominal PDF set the alternative \textsc{MSTW2008nlo68cl}~\cite{mstw2008nnlo}
and CTEQ6.6 PDF sets are investigated.
Events are reweighted according to each of the PDF
uncertainty eigenvectors and the total uncertainty is evaluated 
following the procedure described in Ref.~\cite{top2010}.
An additional uncertainty is assigned for the mis-modelling of jets in the forward $|\eta|$ regions. 
A weight function is derived from the pretag sample by dividing the observed $|\eta|$ distribution 
in data by the distribution obtained from simulated events, for 2-jet and 3-jet events.
The event weights defined in this way are then applied to all simulated samples in the $b$-tagged data set.
The systematic uncertainty is derived from the one-sided difference between the weighted and the nominal samples.
The impact of using simulation samples of limited size is also taken into account.

\vspace{-0.3cm}
\paragraph{Theoretical cross section normalisation} 
The \ttbar, single-top quark $Wt$- and $s$-channel backgrounds are normalised to their theory predictions
with theoretical uncertainties of $^{+7}_{-10}$\%, 
7\% and 4\%, respectively~\cite{Uwer2008,Kidonakis:2010ux,Kidonakis:2010tc}.
The uncertainty on the diboson background is 5\%~\cite{Campbell2011bn}.

\vspace{-0.3cm}
\paragraph{Background normalisation to data} 
The multijet background estimate has an uncertainty of $50\%$. 
The NN analysis places an uncertainty of $50\%$ 
on the rate of events with $W+$heavy flavour jets and $30\%$
on the rate of $W+$light jets events.
These uncertainties are used as constraints on the predictions 
when simultaneously determining the $W+$jets rates and the signal 
cross section.
The cut-based analysis does not apply a global uncertainty on the
$W+$heavy flavour and $W+$light flavour rates, 
but considers separately the impact of the dominant sources of uncertainty
on the data-derived $W+$jets normalisation factors.
This treatment allows the correlation between each 
component of uncertainty on the normalisation factors and the 
uncertainties on the $W+$jets rates to be taken into account.
The $Z$+jets background normalisation has an uncertainty of $60\%$.

\vspace{-0.3cm}
\paragraph{Luminosity} The uncertainty on the integrated luminosity is 
3.7\%~\cite{Aad:2011dr,ATLAS-CONF-2011-116}.\\

Table~\ref{tab:xs-uncertainty-exp} shows the contribution of each source of uncertainty 
to the total uncertainty on the measured $t$-channel cross section 
($\Delta\sigma_\mathrm{obs}/\sigma_\mathrm{obs}$)
for the neural network analysis and for the cut-based analysis.

\begin{table}[!t]
\begin{center}
\caption{Breakdown of the contribution of each source of uncertainty to the 
   total uncertainty of the
   measured $t$-channel cross section in data for the NN analysis and the cut-based analysis.
   Theoretical uncertainties are included in the ``Other backgrounds'' uncertainty category.
}
\label{tab:xs-uncertainty-exp}
  \begin{tabular}{l|c|c}
   \hline
          & \multicolumn{2}{c}{$\Delta\sigma_\mathrm{obs}/\sigma_\mathrm{obs}$ [\%]} \\
   Source                         & NN  &  Cut-based \\
   \hline \hline
    Data statistics               & $\pm$5 & $\pm$8 \\
    \hline			  
    Object modelling             & & \\
    ~~~Jets                       & $\pm$6  & +3/--4  \\  
    ~~~$b$-tagging efficiency     & $\pm$13 & $\pm$12 \\
    ~~~Mistagging rate            & $\pm$1  & $\pm$1 \\
    ~~~Lepton                     & $\pm$2  & $\pm$4 \\   
    ~~~\met, calorimeter readout  & $\pm$2  & $\pm$2 \\   
    \hline
    Monte-Carlo                   &&\\
    ~~~PDF                        & $\pm$3  & $\pm$4 \\
    ~~~Generator                  & $\pm$4  & $\pm$7 \\  
    ~~~Parton shower              & $\pm$5  & $\pm$11 \\
    ~~~ISR/FSR                    & $\pm$14 & +19/--18 \\
    ~~~Forward jet modelling      & +6/--4   & +7/--5   \\  
    ~~~MC statistics              & $\pm$3  & $\pm$4 \\
    \hline
    Background normalisation      &&\\
    ~~~Multijets                  & $\pm$4   & $\pm$2  \\
    ~~~Other backgrounds          & $\pm$1  & $\pm$6 \\
    \hline
    Luminosity                    & $\pm$4  & $\pm$4 \\
    \hline
    \hline
    Total systematic uncertainties           &  +24/--23 & +30/--27 \\
    Total uncertainty                        &  $\pm$24 & +31/--28 \\
    \hline
   \end{tabular}
\end{center}
\end{table}

\section{Cross section measurements}
\label{sec:xs_measure}

Both the cut-based and neural network analyses employ a maximum-likelihood fit method to measure the single top-quark 
$t$-channel cross section.
The general likelihood function is given by the product of the Poisson
likelihoods in the individual channels. The 
background rates are constrained by Gaussian priors.
We use the following equations:

\begin{equation*}
L(\beta^\mathrm{s}; \beta^\mathrm{b}_j) = \prod_{k=1}^M 
\frac{e^{-\mu_k}\cdot\mu_k^{n_k}}{n_k!}\;\cdot\;
\prod_{j=1}^{B} G(\beta^\mathrm{b}_j; 1, \Delta_j) \ \ \ \ \mathrm{with}
\end{equation*}
\begin{equation*}
\mu_k = \mu_k^\mathrm{s} + \sum_{j=1}^{B} \mu_{jk}^\mathrm{b} \ , \ \ \ 
\mu_k^\mathrm{s} = \beta^\mathrm{s}\cdot \tilde{\nu}_\mathrm{s}\cdot\alpha^\mathrm{s}_k \ , 
\ \ \ \mathrm{and} \ \ \
\mu_{jk}^\mathrm{b} = \beta^\mathrm{b}_j\cdot\tilde{\nu}_j\cdot\alpha_{jk}\ ,
\end{equation*}

\noindent where $M$ is the number of channels and $B$ the number of background processes.
The cut-based analysis uses $M=4$ channels separated by lepton charge and the number of
jets, while in the NN-based analysis $M$ is equal to 28, namely the number of bins of the
NN discriminant in the 2-jet channel plus the number of bins of the
NN discriminant in the 3-jet channel.  
Here $\tilde{\nu}_\mathrm{s}$ and $\tilde{\nu}_j$ are, respectively, 
the predicted number of signal events and the 
number of events of background $j$ in the selected data set.
The number of observed (expected) events in channel $k$ is denoted by $n_k$ ($\mu_k$).
The expected number of events in channel $k$ is $\mu^\mathrm{s}_k$ for the signal and 
$\mu^\mathrm{b}_{jk}$
for each background $j$.
The fraction of events falling in channel $k$ is given by 
$\alpha^\mathrm{s}_{k}$ and $\alpha_{jk}$
for signal and background respectively.
For the NN-based analysis the sets of $\alpha^\mathrm{s}_{k}$ and $\alpha_{jk}$
constitute the probability densities (shapes) of the NN discriminants.
The scale factors $\beta^\mathrm{s}$ for signal and $\beta^\mathrm{b}_j$ for the 
backgrounds are the parameters
of the likelihood function that are fitted to the data.
The Gaussian contraints on the background scale factors, $\Delta_j$, are set 
to the theoretical cross section uncertainty for all background processes 
that do not use data-based estimates ($t\bar{t}$, $Wt$, $s$-channel and diboson),
while $W$+jets backgrounds are constrained within their data-derived uncertainties.
The multijet background is fixed to the value estimated from data.

The systematic uncertainties on the cross section measurement are 
determined using a frequentist method based on a large number of pseudo-experiments.
For each pseudo-experiment the expectation values of the backgrounds
$\tilde{\nu}_j$ and of the signal $\tilde{\nu}_\mathrm{s}$ as well as the
relative distribution of events across the channels 
($\alpha^\mathrm{s}_{k}$ and $\alpha_{jk}$) are varied including
all sources of uncertainties described in Section~\ref{sec:systematics}
and the $t$-channel cross section is measured with the maximum-likelihood fit.
The distribution of measured cross sections is an estimator of the probability
density function of all possible outcomes of the measurement and it is 
used to estimate the uncertainty on the actual measurement.

The NN-based analysis yields, from a simultaneous measurement in the 2-jet and 3-jet channels,
a cross section of 
\[ \sigma_{t}= 83\pm 4\,(\mathrm{stat.})\;^{+20}_{-19}\,(\mathrm{syst.})\;\mathrm{pb}
   = 83 \pm 20\;\mathrm{pb.}
\]
The significance of the observed signal corresponds to $7.2$ standard deviations 
($6.0$ expected). This is computed using as a test statistic the $Q$-value,
which is defined as the ratio of the value of the 
likelihood function maximized for the Standard Model signal cross section 
to the value of the likelihood function maximized for zero signal. 

The cut-based analysis measures, by combining four different channels 
(positive and negative lepton charge, with two and three jets)
a cross section of $\sigma_{t}= 92 \,^{+29}_{-26}$~pb,
in good agreement with the NN-based measurement. 
The separation of candidate events according to the lepton charge 
allows individual measurements of the top-quark and top-antiquark
cross sections, yielding the results
$\sigma(t)= 59^{+18}_{-16}\ \mathrm{pb}$ and $\sigma(\bar{t})=33^{+13}_{-12}\ \mathrm{pb,}$
that can be compared to the theoretically predicted cross sections of $41.9^{+1.8}_{-0.8}$~pb 
and $22.7^{+0.9}_{-1.0}$~pb, respectively~\cite{Kidonakis:2011wy}.

To test the compatibility, the two measurements from the NN-based and cut-based analyses 
are combined using the Best Linear Unbiased Estimator (BLUE) method~\cite{Lyons:1988rp}. 
The correlation coefficient of the two analyses is 75\% and was determined
with ensemble tests including all systematic uncertainties.
Based on the ensemble tests the two results are found to be compatible
within one standard deviation.  
However, the combined result and its uncertainty for the observed cross section measurement 
does not significantly differ from the result obtained with the NN analysis alone.

\section{$V_{tb}$ measurement }
\label{sec:Vtb}

Single top-quark production in the $t$-channel proceeds via a $W$-$t$-$b$ vertex and
the measured cross section is proportional to $|V_{tb}|^2$, where
$V_{tb}$ is the relevant CKM matrix element. In the Standard Model $|V_{tb}|$ is close to
one, but new physics contributions could alter its value significantly.

The $|V_{tb}|$ measurement is independent of assumptions about the number of 
quark generations or about the unitarity of the CKM matrix.
The only assumptions required are that $|V_{tb}|\gg|V_{td}|,|V_{ts}|$
and that the $W$-$t$-$b$ interaction is an SM-like left-handed weak coupling.
Therefore, the $t\bar{t}$ background rate is unaffected by a variation 
of $|V_{tb}|$ since decays to a potential higher generation are 
prohibited by kinematics.
On the other hand, rates of single-top quark $Wt$ and $s$-channel backgrounds also scale with $|V_{tb}|^2$,
but their contributions are small in the signal region that drives the maximum-likelihood fit measurement.
The resulting variation on the total top-quark background yield is less than its systematic uncertainty and thus considered negligible.

The value of $|V_{tb}|^2$ is extracted by dividing the
observed single top-quark $t$-channel cross section,
measured using the NN method, by the 
SM expectation~\cite{Kidonakis:2011wy}. 
The experimental and theoretical uncertainties are added in quadrature.
The result obtained is 
$|V_{tb}| = 1.13^{+0.14}_{-0.13}\,(\mathrm{exp.})\pm 0.02\,(\mathrm{theo.}) = 1.13^{+0.14}_{-0.13}$.
Restricting the range of $|V_{tb}|$ to the interval $[0, 1]$, as required by
the SM, a lower limit on $|V_{tb}|$ is extracted: 
$|V_{tb}|>0.75$ at the 95\% confidence level.

\section{Conclusion}
\label{sec:conclusion}

In summary, we present a measurement of the cross section of single top-quark production 
in the $t$-channel with the ATLAS detector in $pp$ collisions at $\sqrt{s}=7$~\tev.
The measurement is based on a neural network discriminant separating signal events
from background and yields a cross section of $83 \pm 20~{\rm pb}$.
The corresponding coupling at the $W$-$t$-$b$ vertex is 
$|V_{tb}|=1.13^{+0.14}_{-0.13}$ and the
95\% confidence level lower limit on the CKM matrix element $|V_{tb}|$ is 0.75.

\section{Acknowledgements}

We thank CERN for the very successful operation of the LHC, as well as the
support staff from our institutions without whom ATLAS could not be
operated efficiently.

We acknowledge the support of ANPCyT, Argentina; YerPhI, Armenia; ARC,
Australia; BMWF, Austria; ANAS, Azerbaijan; SSTC, Belarus; CNPq and FAPESP,
Brazil; NSERC, NRC and CFI, Canada; CERN; CONICYT, Chile; CAS, MOST and NSFC,
China; COLCIENCIAS, Colombia; MSMT CR, MPO CR and VSC CR, Czech Republic;
DNRF, DNSRC and Lundbeck Foundation, Denmark; EPLANET and ERC, European Union;
IN2P3-CNRS, CEA-DSM/IRFU, France; GNAS, Georgia; BMBF, DFG, HGF, MPG and AvH
Foundation, Germany; GSRT, Greece; ISF, MINERVA, GIF, DIP and Benoziyo Center,
Israel; INFN, Italy; MEXT and JSPS, Japan; CNRST, Morocco; FOM and NWO,
Netherlands; RCN, Norway; MNiSW, Poland; GRICES and FCT, Portugal; MERYS
(MECTS), Romania; MES of Russia and ROSATOM, Russian Federation; JINR; MSTD,
Serbia; MSSR, Slovakia; ARRS and MVZT, Slovenia; DST/NRF, South Africa;
MICINN, Spain; SRC and Wallenberg Foundation, Sweden; SER, SNSF and Cantons of
Bern and Geneva, Switzerland; NSC, Taiwan; TAEK, Turkey; STFC, the Royal
Society and Leverhulme Trust, United Kingdom; DOE and NSF, United States of
America.

The crucial computing support from all WLCG partners is acknowledged
gratefully, in particular from CERN and the ATLAS Tier-1 facilities at
TRIUMF (Canada), NDGF (Denmark, Norway, Sweden), CC-IN2P3 (France),
KIT/GridKA (Germany), INFN-CNAF (Italy), NL-T1 (Netherlands), PIC (Spain),
ASGC (Taiwan), RAL (UK) and BNL (USA) and in the Tier-2 facilities
worldwide.

\begin{itshape}

\end{itshape}

\bibliographystyle{model1a-num-names}

\bibliography{finalpaper}

\clearpage
\onecolumn

\begin{flushleft}
{\Large The ATLAS Collaboration}

\bigskip

G.~Aad$^{\rm 48}$,
B.~Abbott$^{\rm 110}$,
J.~Abdallah$^{\rm 11}$,
S.~Abdel~Khalek$^{\rm 114}$,
A.A.~Abdelalim$^{\rm 49}$,
A.~Abdesselam$^{\rm 117}$,
O.~Abdinov$^{\rm 10}$,
B.~Abi$^{\rm 111}$,
M.~Abolins$^{\rm 87}$,
O.S.~AbouZeid$^{\rm 157}$
% ...
\bigskip

% ...
$^{10}$ Institute of Physics, Azerbaijan Academy of Sciences, Baku, Azerbaijan\\
$^{11}$ Institut de F\'isica d'Altes Energies and Departament de F\'isica de la Universitat Aut\`onoma  de Barcelona and ICREA, Barcelona, Spain\\
% ...
$^{48}$ Fakult\"{a}t f\"{u}r Mathematik und Physik, Albert-Ludwigs-Universit\"{a}t, Freiburg i.Br., Germany\\
$^{49}$ Section de Physique, Universit\'e de Gen\`eve, Geneva, Switzerland\\
% ...
$^{87}$ Department of Physics and Astronomy, Michigan State University, East Lansing MI, United States of America\\
% ...
$^{110}$ Homer L. Dodge Department of Physics and Astronomy, University of Oklahoma, Norman OK, United States of America\\
$^{111}$ Department of Physics, Oklahoma State University, Stillwater OK, United States of America\\
% ...
$^{114}$ LAL, Univ. Paris-Sud and CNRS/IN2P3, Orsay, France\\
% ...
$^{117}$ Department of Physics, Oxford University, Oxford, United Kingdom\\
% ...
$^{157}$ Department of Physics, University of Toronto, Toronto ON, Canada\\
% ...
\end{flushleft}

\end{document}